\documentclass[british]{article}
\usepackage[T1]{fontenc}
\usepackage[latin9]{inputenc}
\usepackage[a4paper]{geometry}
\usepackage{color}
\usepackage{babel}
\usepackage{float}
\usepackage{textcomp}
\usepackage{url}
\usepackage{amstext}
\usepackage{graphicx}
\usepackage[authoryear]{natbib}
\usepackage[unicode=true,
 bookmarks=true,bookmarksnumbered=false,bookmarksopen=false,
 breaklinks=true,pdfborder={0 0 0},backref=false,colorlinks=true]
 {hyperref}
\hypersetup{pdftitle={Variation of the Circinus photometry},
 pdfauthor={Konrad R. W. Tristram},
 citecolor = [rgb]{0.0,0.0,0.9}, linkcolor = [rgb]{0.0,0.0,0.9}, urlcolor  = [rgb]{0.0,0.0,0.9}}

\makeatletter

\providecommand{\tabularnewline}{\\}

\usepackage[normal,normalsize,bf]{caption}
\usepackage[bf,small]{titlesec}
\date{October 12, 2009}

\let\jnl@style=\rmfamily
\def\ref@jnl#1{{\jnl@style#1}}%
\newcommand\aj{\ref@jnl{AJ}}%
\newcommand\araa{\ref@jnl{ARA\&A}}%
\newcommand\apj{\ref@jnl{ApJ}}%
\newcommand\apjl{\ref@jnl{ApJ}}%
\newcommand\apjs{\ref@jnl{ApJS}}%
\newcommand\apss{\ref@jnl{Ap\&SS}}%
\newcommand\aap{\ref@jnl{A\&A}}%
\newcommand\aapr{\ref@jnl{A\&A~Rev.}}%
\newcommand\aaps{\ref@jnl{A\&AS}}%
\newcommand\mnras{\ref@jnl{MNRAS}}%
\newcommand\nar{\ref@jnl{New A Rev.}}%
\newcommand\pasp{\ref@jnl{PASP}}%
\newcommand\nat{\ref@jnl{Nature}}%
\newcommand\procspie{\ref@jnl{Proc.~SPIE}}%
\makeatother

\begin{document}

\title{\vspace{-1.5cm}
Stability of the MIDI photometry: the case of Circinus}

\author{Konrad R. W. Tristram (\href{mailto:tristram@mpifr-bonn.mpg.de}{tristram@mpifr-bonn.mpg.de})}

\maketitle
In principle, MIDI should always measure the same calibrated total
flux spectrum for a specific source, independent of the instrument
settings and the baseline geometry. In the data on the Circinus galaxy,
however, there is (a) a general \textit{offset} of the flux values
for 2009 and (b) a slow \textit{drift} of the total fluxes at short
wavelengths during two nights (2008-04-17 and 2009-04-14). The latter
seems to depend on the hour angle of the observation. The two distinct
variations will be referred to as \textit{``offset}'' and ``\textit{drift}''
in the following. In this document, a more detailed analysis of these
two effects is carried out and summarised. The goal is to find an
explanation for these variations in the photometry.

\begin{figure}[H]
\vspace{0.5cm}

\begin{centering}
\includegraphics[viewport=140bp 296bp 467bp 552bp,clip,width=0.49\textwidth]{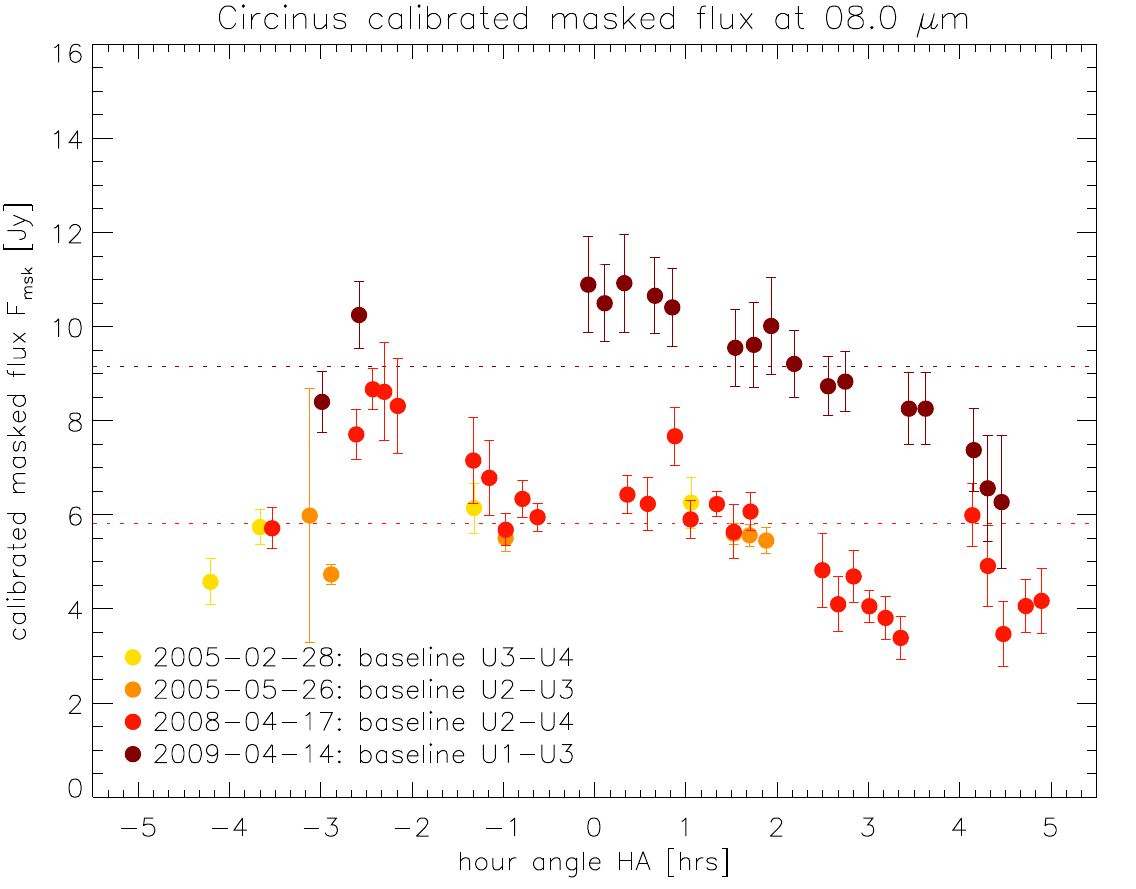}\hfill{}\includegraphics[viewport=135bp 296bp 467bp 552bp,clip,width=0.49\textwidth]{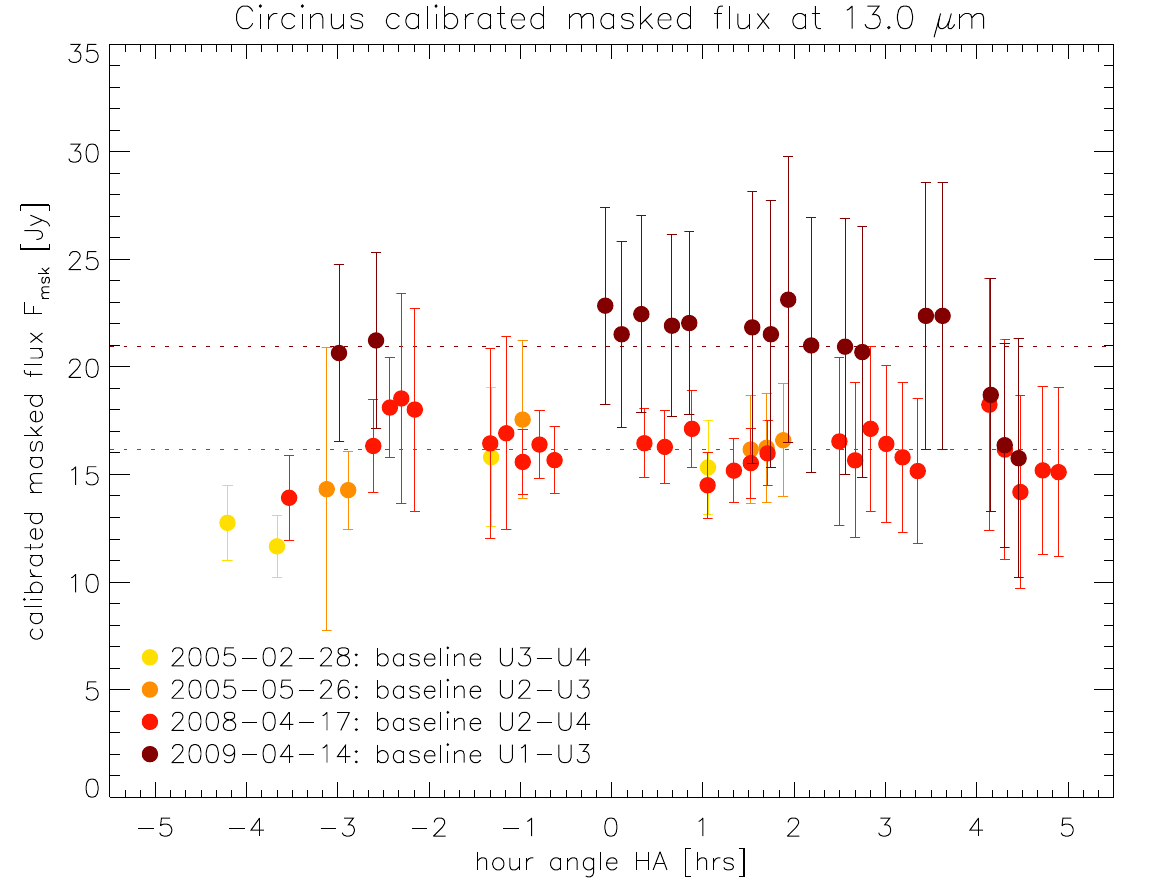}
\par\end{centering}

\vspace{0.5cm}

\begin{centering}
\includegraphics[viewport=140bp 296bp 467bp 552bp,clip,width=0.49\textwidth]{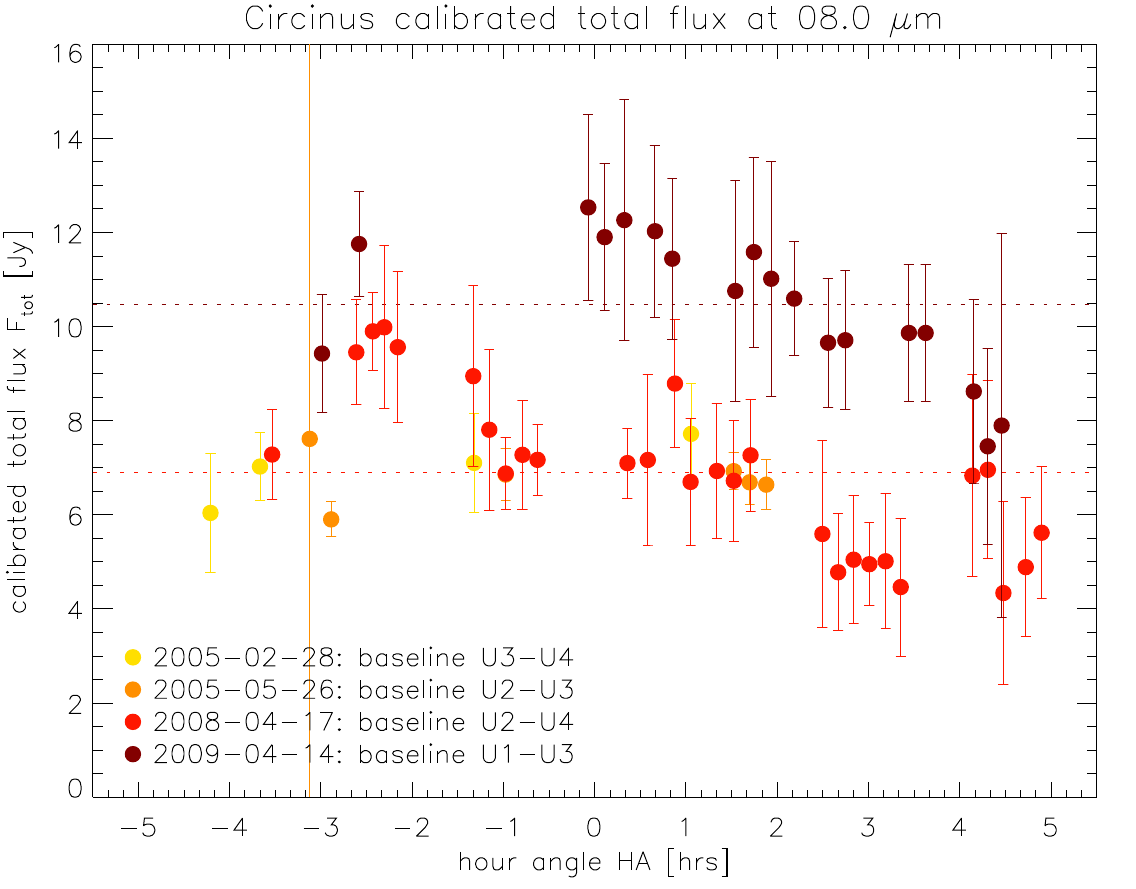}\hfill{}\includegraphics[viewport=135bp 296bp 467bp 552bp,clip,width=0.49\textwidth]{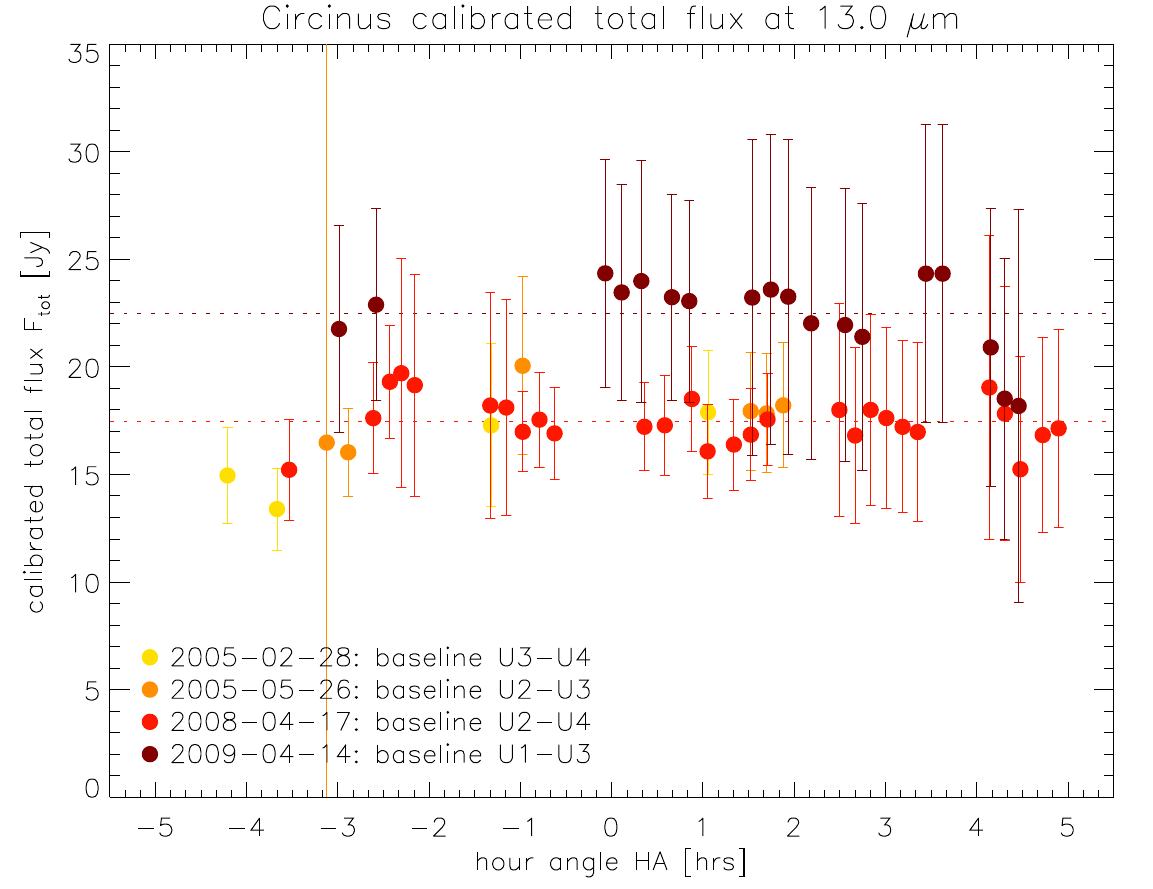}
\par\end{centering}

\caption{Masked (top row) and total (bottom row) flux at $8$ and $10\,\mathrm{\mu m}$
as a function of the hour angle, $\mathrm{HA}$, for four different
baselines and dates. The average flux values in 2008-04-17 (U2-U4)
and 2009-04-14 (U1-U3) are indicated by the dashed lines.\label{fig:fluxes_hourangle}}
\end{figure}

\begin{figure}[p]
\begin{centering}
\includegraphics[viewport=140bp 296bp 467bp 552bp,clip,width=0.49\textwidth]{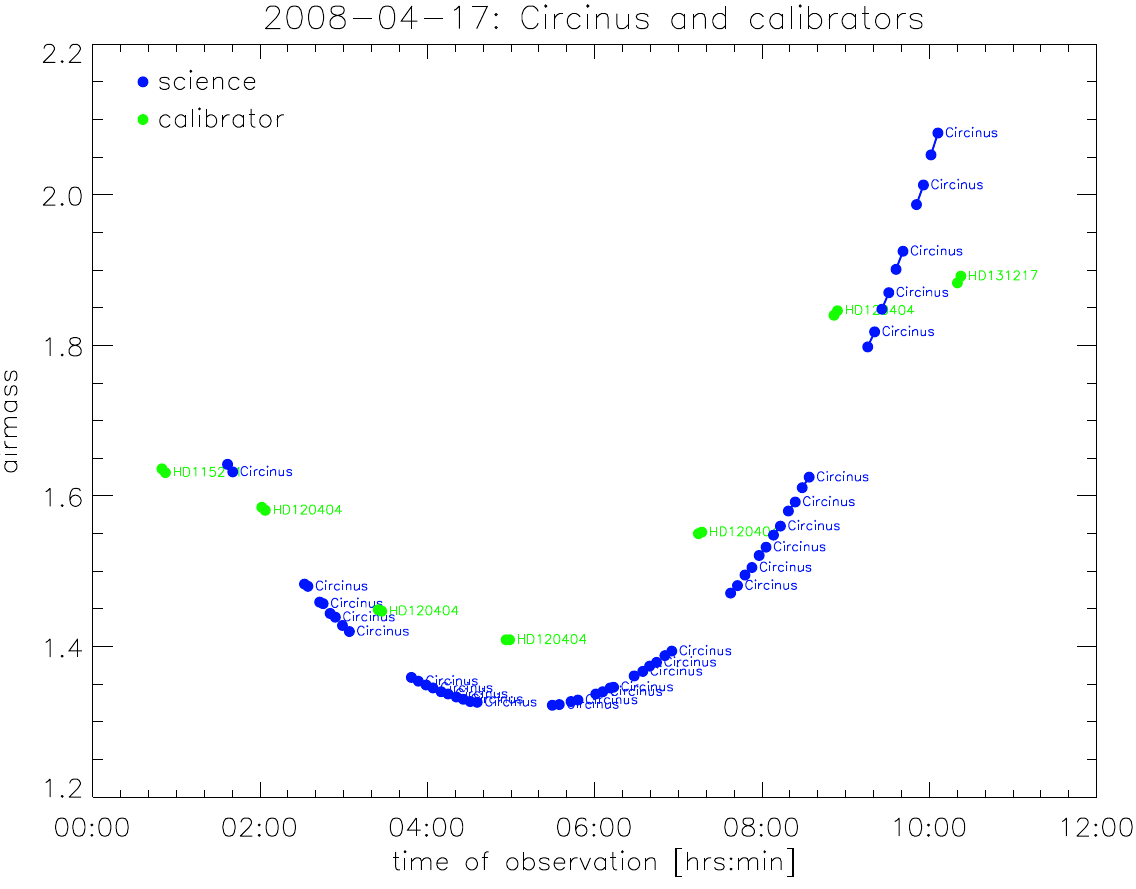}\hfill{}\includegraphics[viewport=135bp 296bp 467bp 552bp,clip,width=0.49\textwidth]{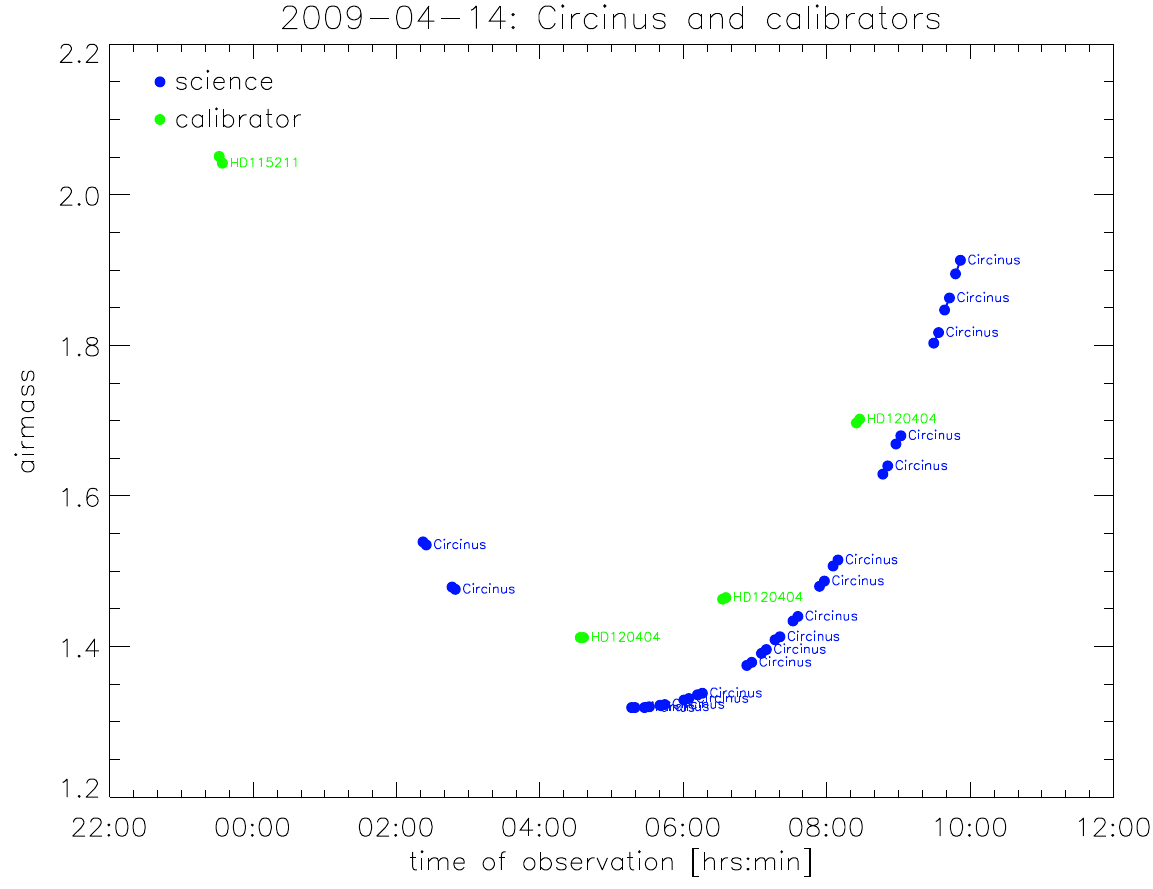}
\par\end{centering}

\vspace{1.5cm}

\begin{centering}
\includegraphics[viewport=140bp 296bp 467bp 552bp,clip,width=0.49\textwidth]{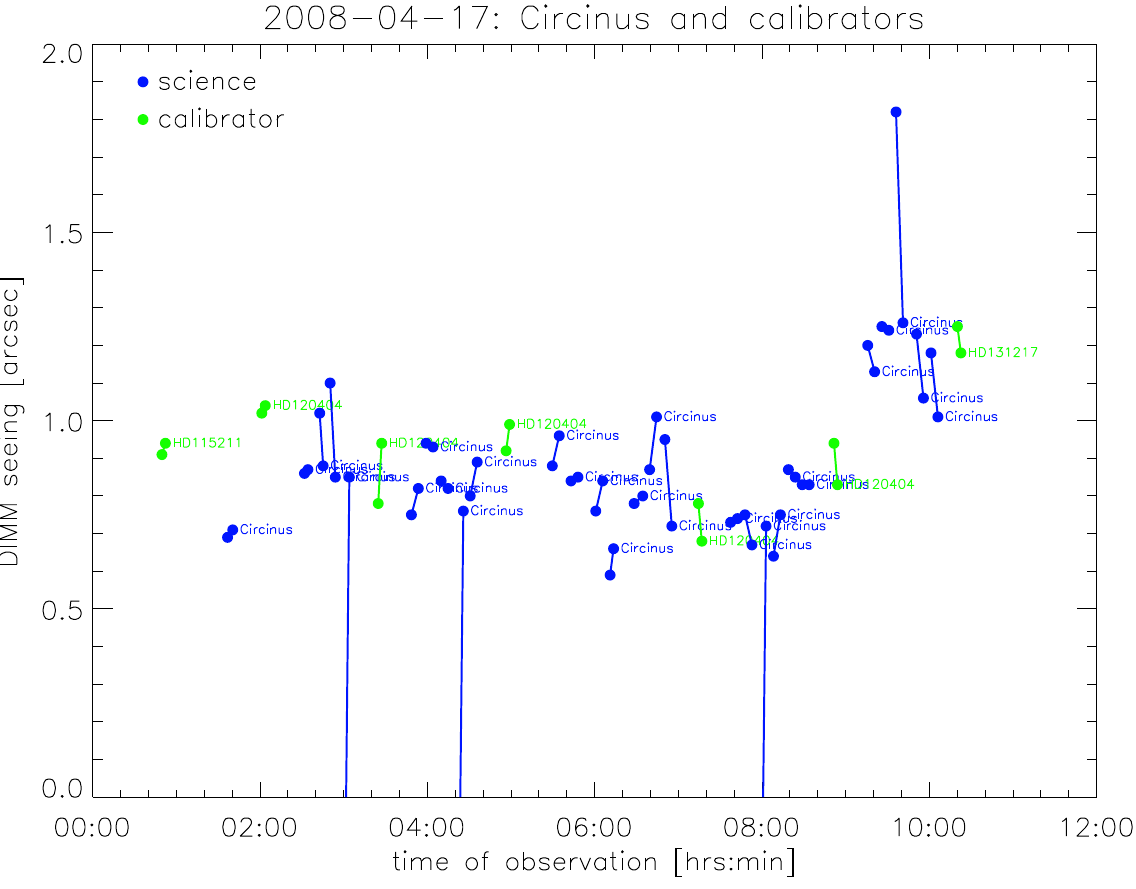}\hfill{}\includegraphics[viewport=135bp 296bp 467bp 552bp,clip,width=0.49\textwidth]{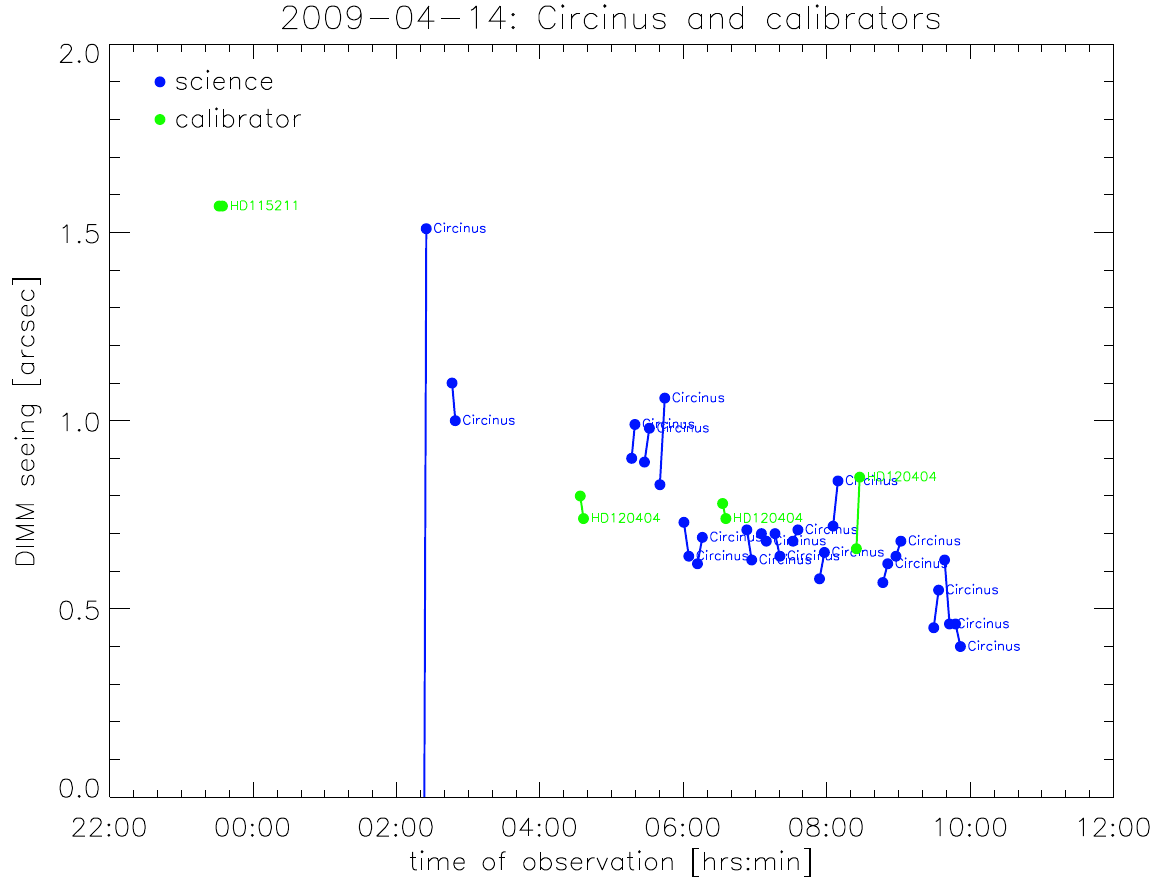}
\par\end{centering}

\vspace{1.5cm}

\begin{centering}
\includegraphics[viewport=140bp 296bp 467bp 552bp,clip,width=0.49\textwidth]{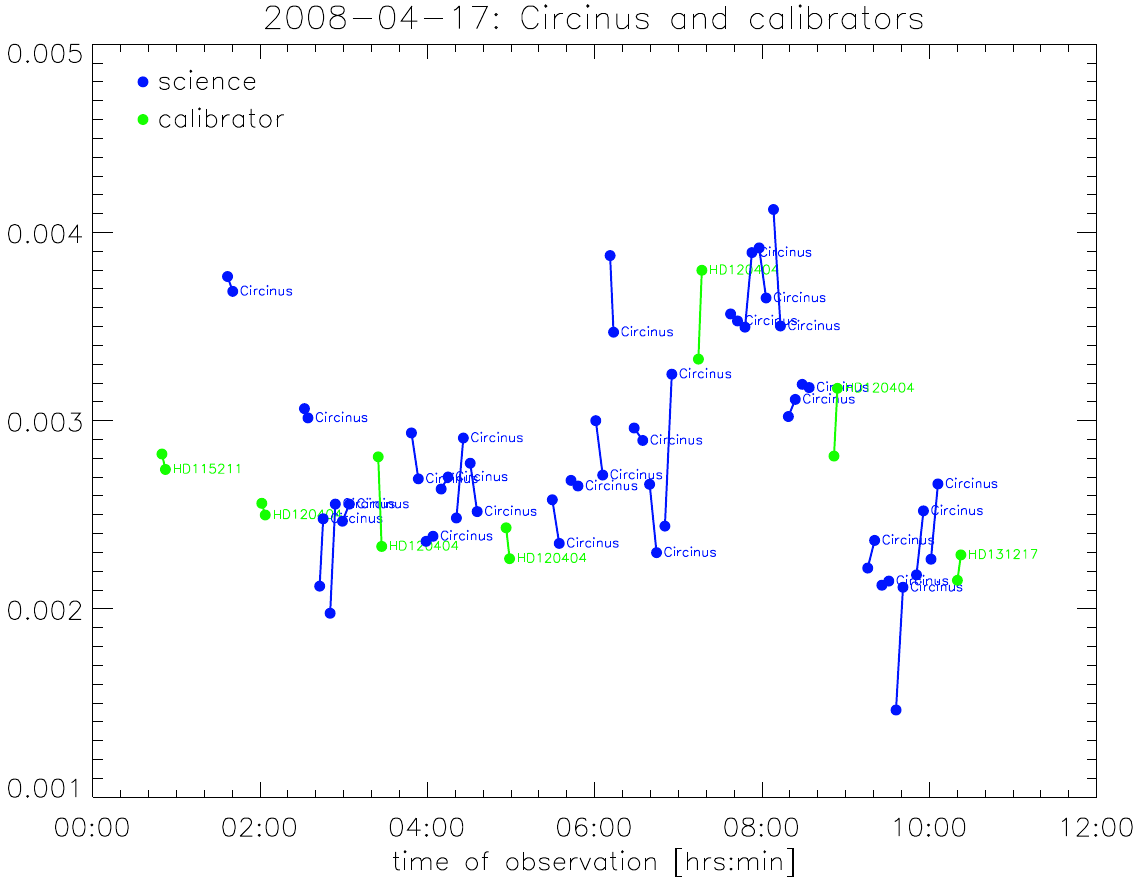}\hfill{}\includegraphics[viewport=135bp 296bp 467bp 552bp,clip,width=0.49\textwidth]{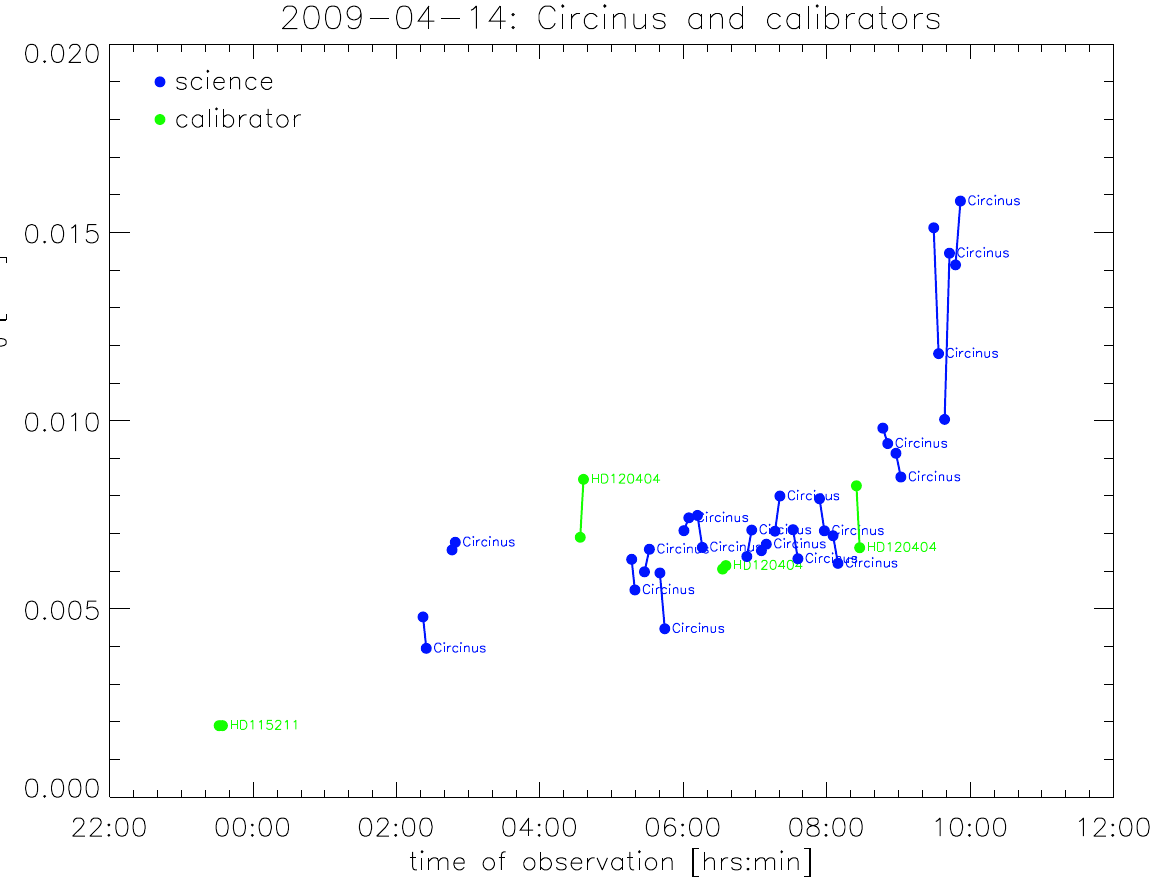}
\par\end{centering}

\caption{Ambient conditions for the measurements on 2008-04-17 (left column)
and on 2009-04-14 (right column) as stored in the FITS headers of
the interferometric data files: airmass (first row), DIMM seeing (second
row) and coherence time (bottom row).\label{fig:ambientheader}}
\end{figure}

For the analysis, I mainly used four data sets obtained on four different
nights, all using different baselines. I looked at both the unmasked
total flux spectra (unmasked photometry, $F_{\mathrm{tot}}$) and
at the masked total flux spectra (masked photometry, $F_{\mathrm{msk}}$),
hereafter ``total flux'' and ``masked flux''. The masked and total
fluxes at two wavelengths, $8$ and $13\,\mathrm{\mu m}$, are shown
in Figure~\ref{fig:fluxes_hourangle}. For 2008-04-17 and 2009-04-14,
the ambient conditions (airmass, DIMM seeing and coherence time) and
the timing of the observations are shown in Figure~\ref{fig:ambientheader}.
A list of the data sets and the corresponding apertures (at $10\,\mathrm{\mu m}$)
within which the spectra were extracted are given in Table~\ref{tab:properties}.
\begin{table}
\begin{centering}
\begin{minipage}[t]{1\columnwidth}%
\renewcommand{\footnoterule}{}

\begin{center}
\begin{tabular}{ccccll}
\hline 
date & baseline & aperture total flux & aperture masked flux & DET DIT & CHOP FREQ\tabularnewline
\hline 
2005-02-28 & U3-U4 & $0.52\times1.38\,\mathrm{arcsec}^{2}$ & $0.52\times0.65\,\mathrm{arcsec}^{2}$ & ~~$12\,\mathrm{ms}$ & ~~$2.0\,\mathrm{Hz}$\tabularnewline
2005-05-26 & U2-U3 & $0.52\times1.38\,\mathrm{arcsec}^{2}$ & $0.52\times0.57\,\mathrm{arcsec}^{2}$ & ~~$18\,\mathrm{ms}$%
\footnote{The first calibrator and Circinus combination uses $12\,\mathrm{ms}$
integration times.%
} & ~~$2.0\,\mathrm{Hz}$\tabularnewline
2008-04-17 & U2-U4 & $0.52\times1.29\,\mathrm{arcsec}^{2}$ & $0.52\times0.60\,\mathrm{arcsec}^{2}$ & ~~$18\,\mathrm{ms}$ & ~~$1.3\,\mathrm{Hz}$%
\footnote{One observation of the Circinus nucleus has $0.5\,\mathrm{Hz}$ chopping.%
}\tabularnewline
2009-04-14 & U1-U3 & $0.52\times1.29\,\mathrm{arcsec}^{2}$ & $0.52\times0.56\,\mathrm{arcsec}^{2}$ & ~~$20\,\mathrm{ms}$ & ~~$0.5\,\mathrm{Hz}$\tabularnewline
\hline 
\end{tabular}
\par\end{center}%
\end{minipage}
\par\end{centering}

\caption{Properties of the photometry data used for the analysis.\label{tab:properties}}

\end{table}

The most obvious difference in the measurements is that the flux values
obtained on 2009-04-14 are in general significantly higher (by $3$
to $5\:\mathrm{Jy}$, that is by $25$ to $50\,\%$, see dashed lines
in Figure~\ref{fig:fluxes_hourangle}) than the values on the other
dates, both for the total and the masked fluxes. This \textit{offset}
is present at all wavelengths, but it is stronger at short wavelengths
relative to the absolute flux values. The fluxes on the other dates
are all consistent with each other.

Although mainly focussed on the correlated fluxes, the higher flux
values measured on 2009-04-14 were already discussed in §2 of my memo
from 2009/06/09. There it turned out, that not only the masked and
total fluxes but also the correlated fluxes measured on 2009-04-14
were higher than those measured on 2008-04-17 (although by far not
as much as the total fluxes). The raw correlated fluxes on 2008-04-17
and 2009-04-14 actually agree very well and it was concluded that
the calibration of the data may be part of the problem. For the masked
and the total fluxes, on the other hand, there are considerable differences
in the raw counts (see §2 in the memo from 2009/06/09), indicating
that the calibration is not the main reason for the increase of the
masked and total flux. In §4 of the same memo, measurements obtained
on 2005-02-28 (yellow point at $\mathrm{HA}=\textrm{1:00}$ in Figure~\ref{fig:fluxes_hourangle})
and 2009-04-17 (brown points at $\mathrm{HA}\sim\textrm{4:30}$ in
Figure~\ref{fig:fluxes_hourangle}) were compared: both the total
and the correlated fluxes agree well, only the differential phases
differ.

The second dependence is the \textit{drift} of the total flux at short
wavelengths with the hour angle on 2008-04-17 and 2009-04-14. At $8\,\mathrm{\mu m}$,
both the masked and the total flux decrease by almost a factor 2 over
the course of the night. Both the total and masked fluxes seem to
have a maximum at $\mathrm{HA}\sim-2.5\,\mathrm{h}$ and a minimum
at $\mathrm{HA}\sim+4.5\,\mathrm{h}$ (as far as this can be judged
from the limited range of hour angles covered, $-4.0\,\mathrm{h}<\mathrm{HA}<+5\,\mathrm{h}$).
This drift is only present for $8\,\mathrm{\mu m}<\lambda<9\,\mathrm{\mu m}$;
there is no apparent variation of the fluxes at long wavelengths,
that is, the spectral shape of the source varies changes during the
night.

In the remainder of this document, different possible explanations
for these variations are discussed.

\section{Instrument settings (drift and offset)}

For the observations all instrument settings (except the number of
frames, NDIT) were generally kept the same between the respective
calibrator and the science data, so that all influences by these settings
should be calibrated away. None of the parameters varied continuously
during an observation epoch, so the drift at short wavelengths must
have been caused by something else. There are, however, a few changes
between the epochs concerning the detector integration times (DET
DIT) and the chopping frequencies (CHOP FREQ). For a detailed listing
of these two parameters, see Table~\ref{tab:properties}. In principle,
this could have led to the offset of the flux in 2009. But because
the respective settings were applied equally to the calibrators and
the Circinus nucleus and it is not explainable why only the raw count
rates of Circinus changed and not those of the calibrators (see Figure~\ref{fig:relcounts}
in Section~\ref{sec:calibrator}).\textbf{ I therefore see no obvious
connection between the changed integration time or chopping frequency
and the drift or offset in the fluxes.}

\section{Airmass (drift)\label{sec:airmass}}

The drift in the flux values during one night could be an effect of
the airmass, as the airmass through which Circinus is being observed
changes during the night. However, the minimum airmass during culmination
of the object is reached at $\mathrm{HA}=0\,\mathrm{h}$. Any variation
of the measured flux should be symmetric to this angle. But this is
not the case in a convincing way. Especially the measurements from
2008 show a continuously decreasing flux from $\mathrm{HA}\sim-2\,\mathrm{h}$
to $\mathrm{HA}\sim3\,\mathrm{h}$ without any symmetry with respect
to $\mathrm{HA}=0\,\mathrm{h}$. Unfortunately, not many observations
were carried out with $\mathrm{HA}<0\,\mathrm{h}$ on 2009-04-14 due
to technical problems.

Also, any such variation in the measured fluxes should have been removed
by the calibration, because always a calibrator at a similar airmass
was used. Finally, the calibrators don't show any significant dependence
on the airmass themselves (see Section~\ref{sec:calibrator} and
Figure~\ref{fig:transfer}). So this should be also true for the
Circinus galaxy, especially because the flux levels are on the order
of $\sim10\,\mathrm{Jy}$ for both the calibrators and the galaxy.
\textbf{I conclude that the airmass is probably not the main reason
for the variation of the masked and the total flux.}

\section{Atmospheric conditions and AO correction (offset and drift)\label{sec:atmosphere}}

The atmospheric conditions and the quality of the AO correction could
lead to both the offset as well as the drift in the total flux, because
the PSF size is wavelength dependent and because the atmospheric conditions
could have changed smoothly in each night and varied strongly between
observing epochs. Figure~\ref{fig:ambientheader} shows that on 2008-04-17
the DIMM seeing varied between $0.6''$ and $1.3''$ with no general
trend over the night (except for an increase at the end of the night).
On 2009-04-14 however, the seeing steadily \textbf{improved} from
$1.5''$ to $0.4''$ over the course of the night, while the measured
flux values \textbf{decreased}. The seeing (as measured by the DIMM)
is thus not the reason for the decrease of the flux over the night.
The average seeing values and their standard deviation for the two
nights are $0.90''\pm0.23''$ and $0.75''\pm0.26''$ for 2008-04-17
and 2009-04-14 respectively.

\begin{figure}
\begin{centering}
\includegraphics[viewport=140bp 296bp 467bp 552bp,clip,scale=0.73]{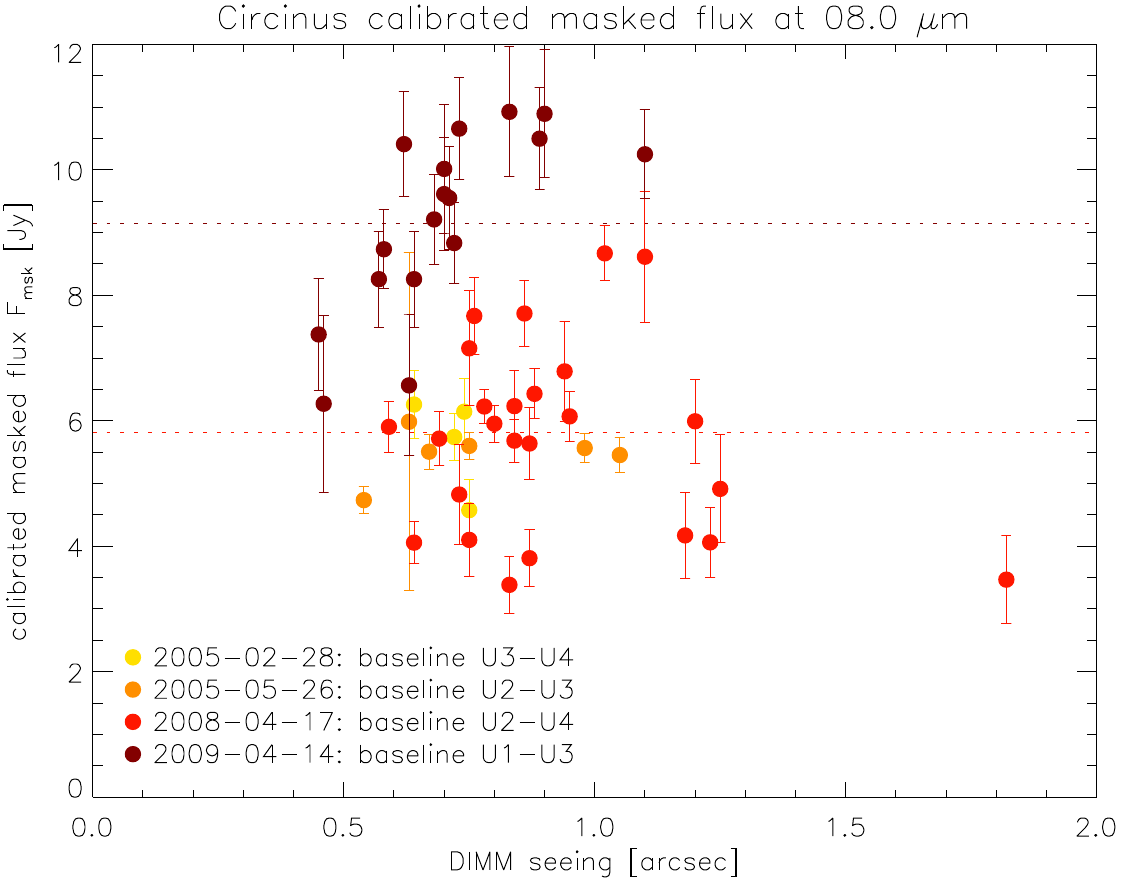}\includegraphics[viewport=135bp 296bp 467bp 552bp,clip,scale=0.73]{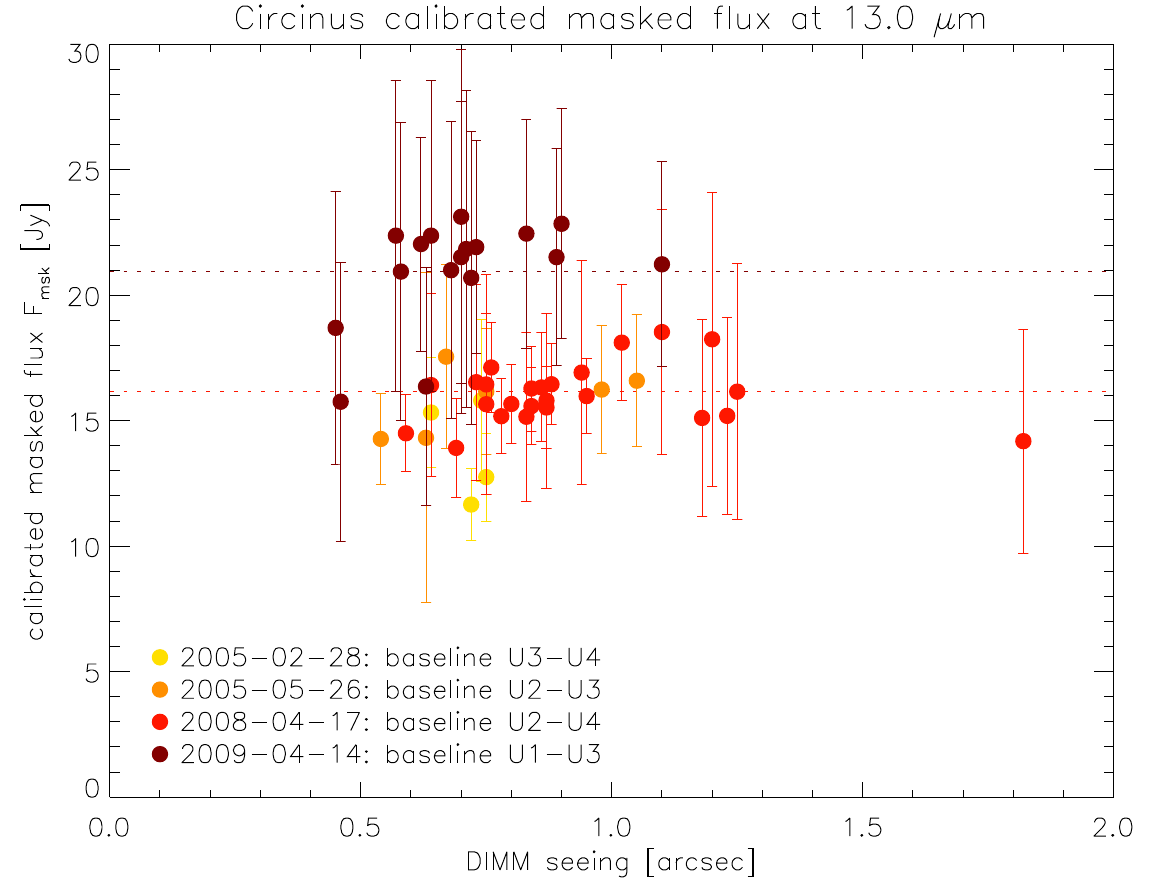}
\par\end{centering}

\caption{Masked flux as a function of the DIMM seeing as stored in FITS header.\label{fig:seeing}}
\end{figure}

Whether the seeing is causing the variations can be more directly
tested by plotting the measured fluxes as a function of the seeing,
see Figure~\ref{fig:seeing}. For worse seeing conditions, one would
expect a worse AO correction and hence a broader PSF which in turn
could lead to a decrease of the measured fluxes. This is however not
the case. The data from 2009-04-14 actually suggests the opposite
trend, higher flux values for worse seeing conditions. \textbf{The
seeing conditions are hence not the reason for the change of the measured
fluxes in Circinus.} The values of the coherence time, $\tau_{0}$,
show more or less the inverse behaviour than the seeing trends and
hence are not responsible for the changes in the measured fluxes either.

\begin{figure}
\begin{centering}
\includegraphics[viewport=140bp 296bp 467bp 552bp,clip,width=0.49\textwidth]{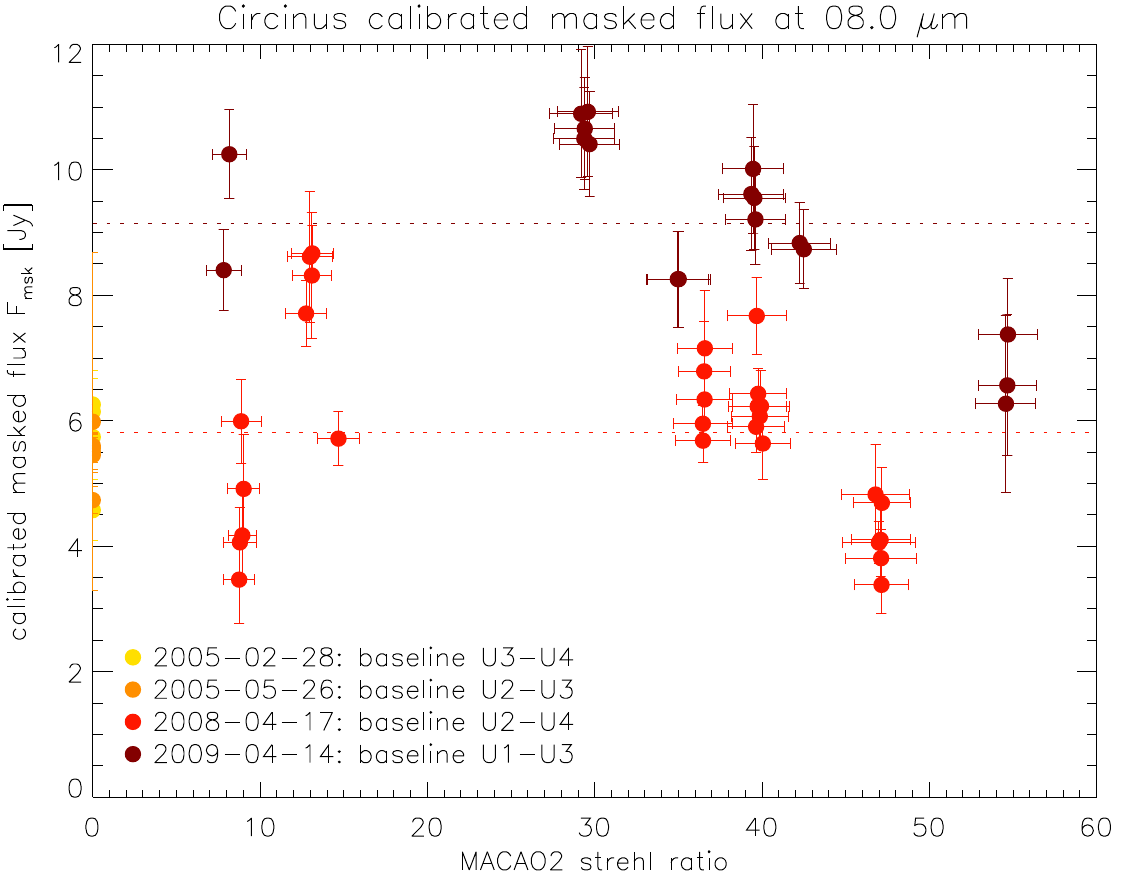}\hfill{}\includegraphics[viewport=135bp 296bp 467bp 552bp,clip,width=0.49\textwidth]{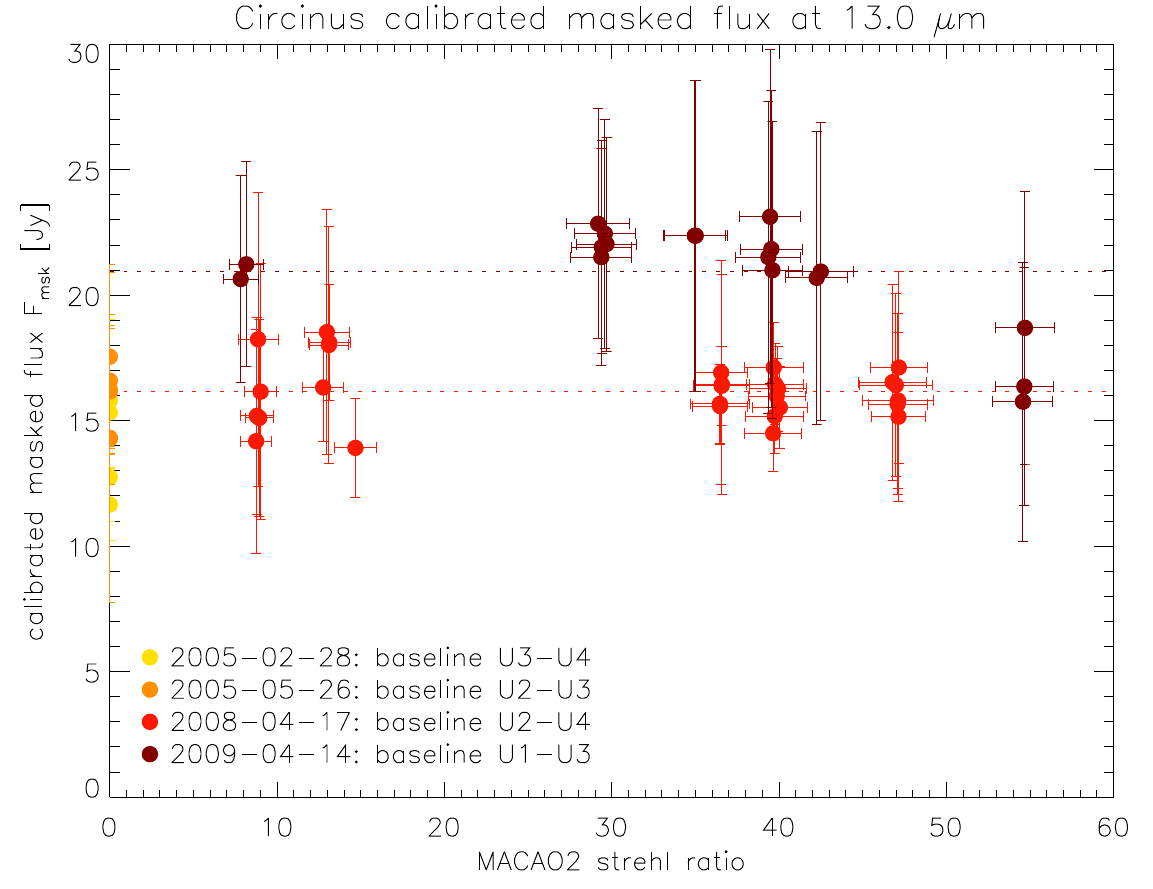}
\par\end{centering}

\caption{Masked flux as a function of the strehl ratio from the MACAO unit
in beam 2. Note that for all our measurements before 2006, all AO
related values in the FITS headers are 0 - probably these were not
stored correctly. \label{fig:strehl}}
\end{figure}

Several parameters of the AO systems (e.g. ``encircled energy'',
``delivered FWHM'' or ``strehl'') roughly correlate with the seeing
or the coherence time. The measured fluxes do not show any clear dependency
on these parameters either as can be seen exemplarily for the strehl
ratio in Figure~\ref{fig:strehl}. However, the values stored in
the AO related FITS header keywords should not be taken too serious
anyway, as far as I have been told.

\section{PSF size (drift and offset)\label{sec:PSFsize}}

\begin{figure}
\begin{centering}
\includegraphics[viewport=140bp 296bp 467bp 552bp,clip,width=0.49\textwidth]{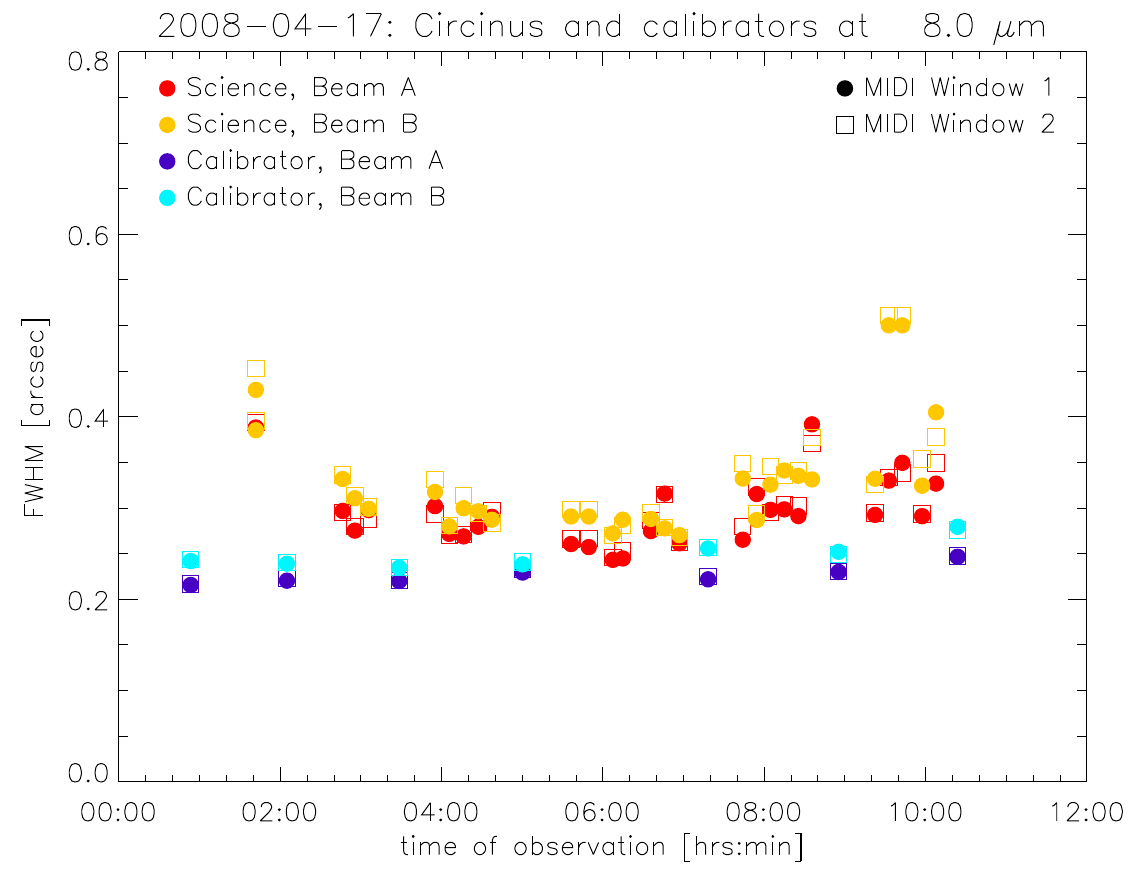}\hfill{}\includegraphics[viewport=140bp 296bp 467bp 552bp,clip,width=0.49\textwidth]{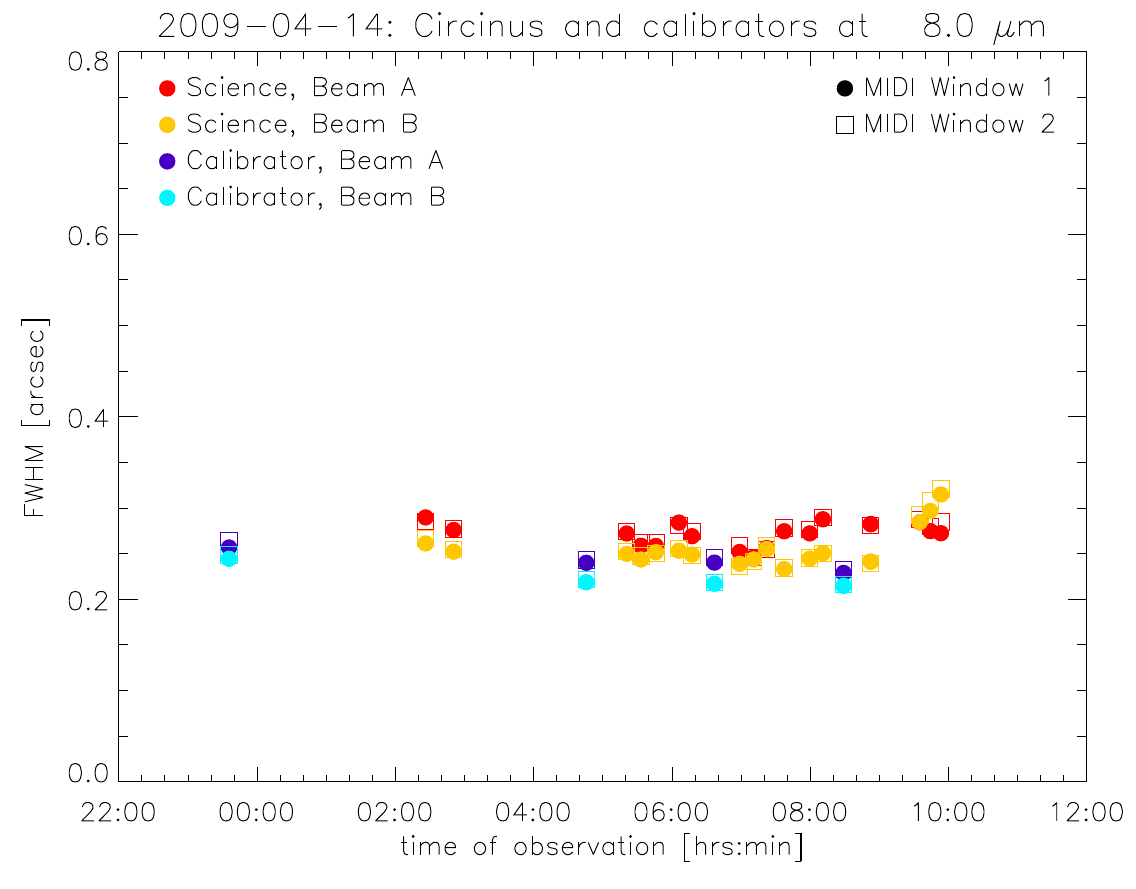}
\par\end{centering}

\begin{centering}
\includegraphics[viewport=140bp 296bp 467bp 552bp,clip,width=0.49\textwidth]{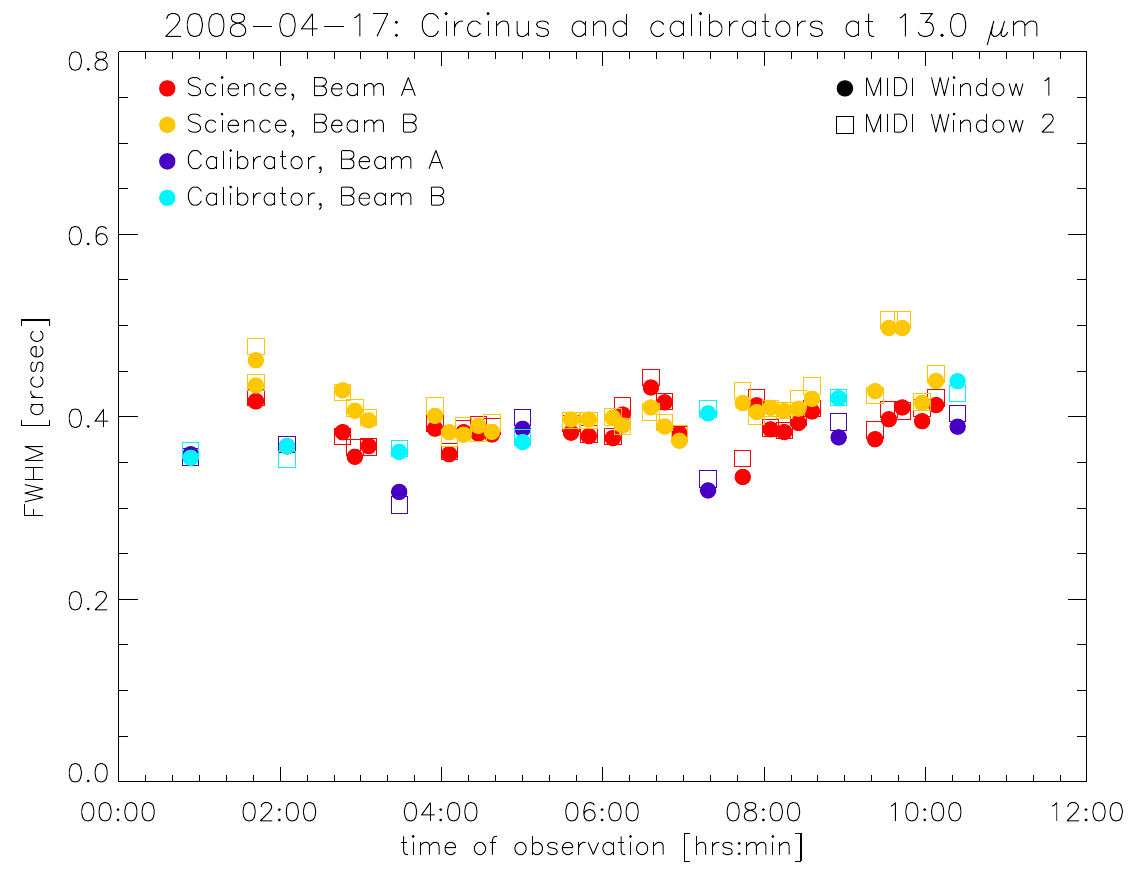}\hfill{}\includegraphics[viewport=140bp 296bp 467bp 552bp,clip,width=0.49\textwidth]{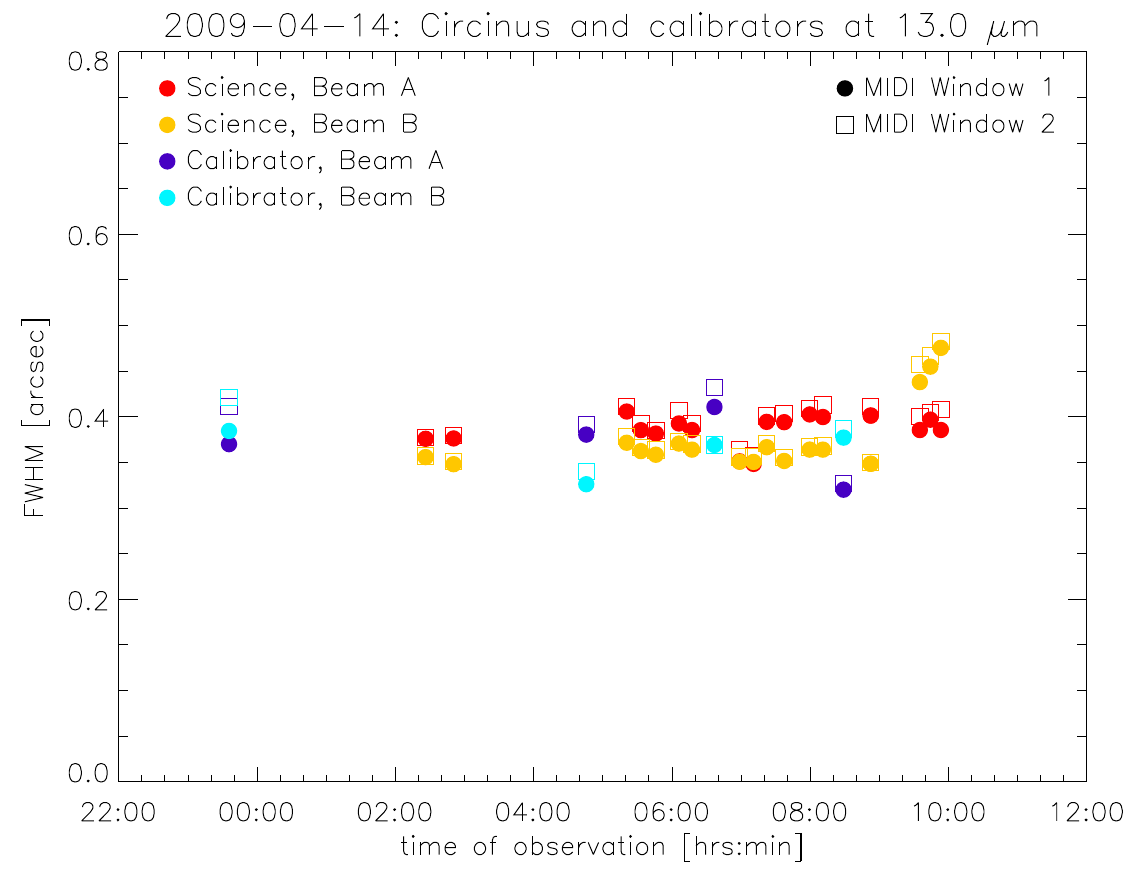}
\par\end{centering}

\caption{FWHM of the spectrum at $8\,\mathrm{\mu m}$ (top row) and at $10\,\mathrm{\mu m}$
(bottom row) for the photometries observed on 2008-04-17 (left column)
and 2009-04-14 (right column). The FWHM are plotted for both the Circinus
galaxy (``Science'') and the calibrators (``Calibrator'') with
different colours for the different beams (i.e. telescopes) and different
symbols for the the different windows in MIDI. \label{fig:fwhm-plots}}
\end{figure}

The size of the PSF can have an influence on the measured flux values,
e.g. by changes in the losses induced by the slit and the mask. To
analyse the PSF quality, I fitted the spatial profiles of all total
flux spectra observed on 2008-04-17 and 2009-04-14 by a Gaussian distribution.
The fit was carried out at two wavelengths: at $8$ and at $13\,\mathrm{\mu m}$.
The FWHM of these Gaussians as a function of the observing time are
shown in Figure~\ref{fig:fwhm-plots}.
 Except for a few slightly broader spectra at $8\,\mathrm{\mu m}$
on 2008-04-17 all spectra have very similar FWHM. Above all, no continuous
drifts in a single night nor significant offsets between the two different
epochs are visible. \textbf{I therefore conclude that the PSF size
did not change much during the observations and that it cannot be
responsible for the observed drift and offset in the fluxes.}

The values of the FWHM are nevertheless interesting in themselves.
At $8\,\mathrm{\mu m}$, the PSF of Circinus is slightly larger than
that of the calibrator star, while at $13\,\mathrm{\mu m}$ both PSF
have comparable sizes (see Table~\ref{tab:fwhm-values}).
\begin{table}
\begin{tabular}{cc@{ }r@{ }r@{ }l@{\extracolsep{1pt}}rc@{ }r@{ }r@{ }l}
\multicolumn{10}{c}{\textbf{2008-04-17}}\tabularnewline
\hline 
object & \multicolumn{4}{c}{FWHM($8\,\mathrm{\mu m}$)} &  & \multicolumn{4}{c}{FWHM($13\,\mathrm{\mu m}$)}\tabularnewline
\hline 
Circinus & $317$ & $\pm$ & $55$ & mas &  & $404$ & $\pm$ & $30$ & mas\tabularnewline
Calibrators & $238$ & $\pm$ & $17$ & mas &  & $375$ & $\pm$ & $34$ & mas\tabularnewline
\hline 
\end{tabular}\hfill{}%
\begin{tabular}{cc@{ }r@{ }r@{ }l@{\extracolsep{1pt}}rc@{ }r@{ }r@{ }l}
\multicolumn{10}{c}{\textbf{2009-04-14}}\tabularnewline
\hline 
object & \multicolumn{4}{c}{FWHM($8\,\mathrm{\mu m}$)} &  & \multicolumn{4}{c}{FWHM($13\,\mathrm{\mu m}$)}\tabularnewline
\hline 
Circinus & $265$ & $\pm$ & $20$ & mas &  & $384$ & $\pm$ & $32$ & mas\tabularnewline
Calibrators & $235$ & $\pm$ & $16$ & mas &  & $376$ & $\pm$ & $35$ & mas\tabularnewline
\hline 
\end{tabular}

\caption{Average FWHM and standard deviations of the FWHM of the spectra measured
on 2008-04-17 and 2009-04-14. The FWHM of the diffraction limited
PSFs of an $8.2\,\mathrm{m}$ telescope at $8$ and $13\,\mathrm{\mu m}$
are $\sim210$ and $\sim340\,\mathrm{mas}$, respectively.\label{tab:fwhm-values}}
\end{table}
 This means that, in the MIR at $8\,\mathrm{\mu m}$, Circinus is
already slightly resolved by an AO assisted $8\mathrm{m}$-class telescope.
In both nights the FWHM of the PSF of the Circinus nucleus seems to
have been slightly broader at the beginning and at the end of the
night, where observations were carried out at higher airmasses (c.f.
Figure~\ref{fig:ambientheader}). In 2009 the effect is only visible
for the end of the night because observations of Circinus only started
only in the middle of the night due to technical problems earlier.
On 2009-04-14 the atmospheric conditions improved significantly over
the course of the night (see Figure~\ref{fig:ambientheader}), the
PSF size in the MIR did, on the other hand, not change much. This
indicates that the AO correction provided a relatively stable PSF
in the MIR, irrespective of the atmospheric conditions (see also Figure
10 in my memo from 2008/05/29).

\section{Variations in the calibrator data (drift and offset)\label{sec:calibrator}}

\begin{figure}
\begin{centering}
\includegraphics[viewport=137bp 296bp 467bp 552bp,clip,width=0.49\textwidth]{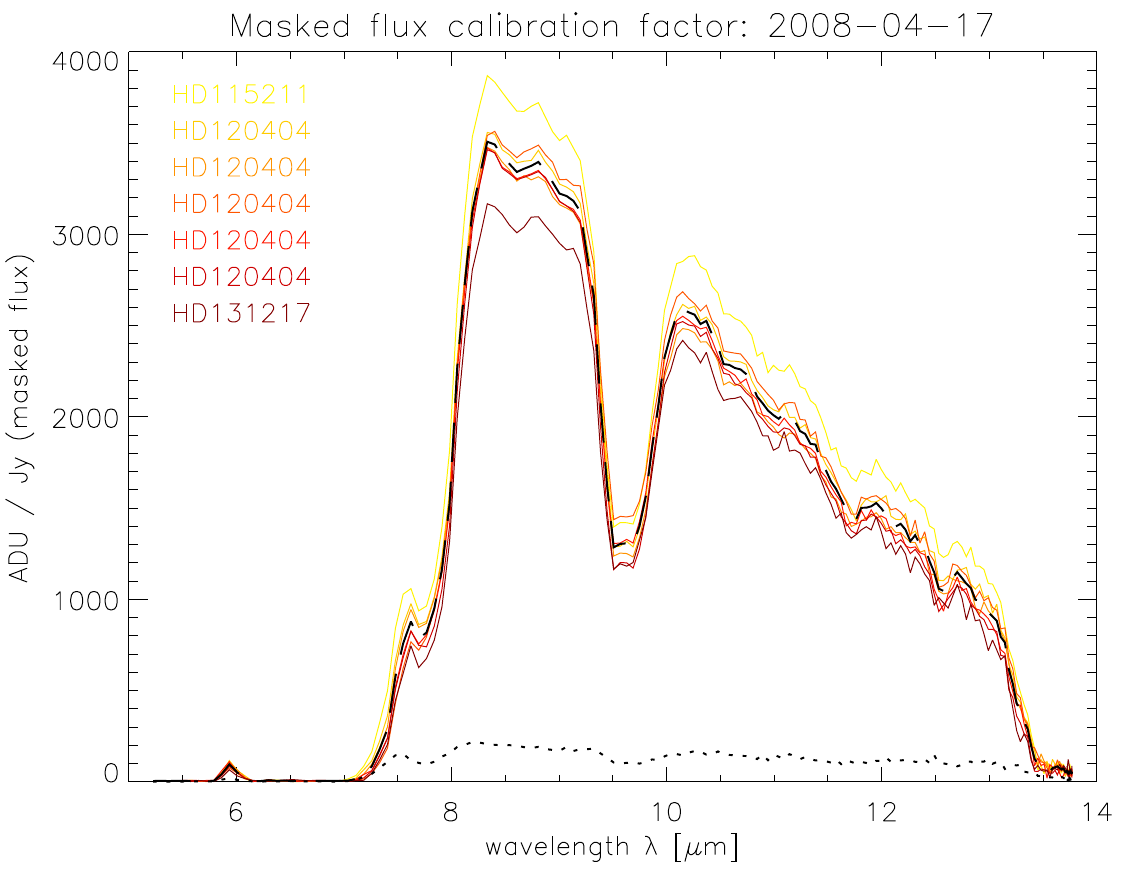}\hfill{}\includegraphics[viewport=137bp 296bp 467bp 552bp,clip,width=0.49\textwidth]{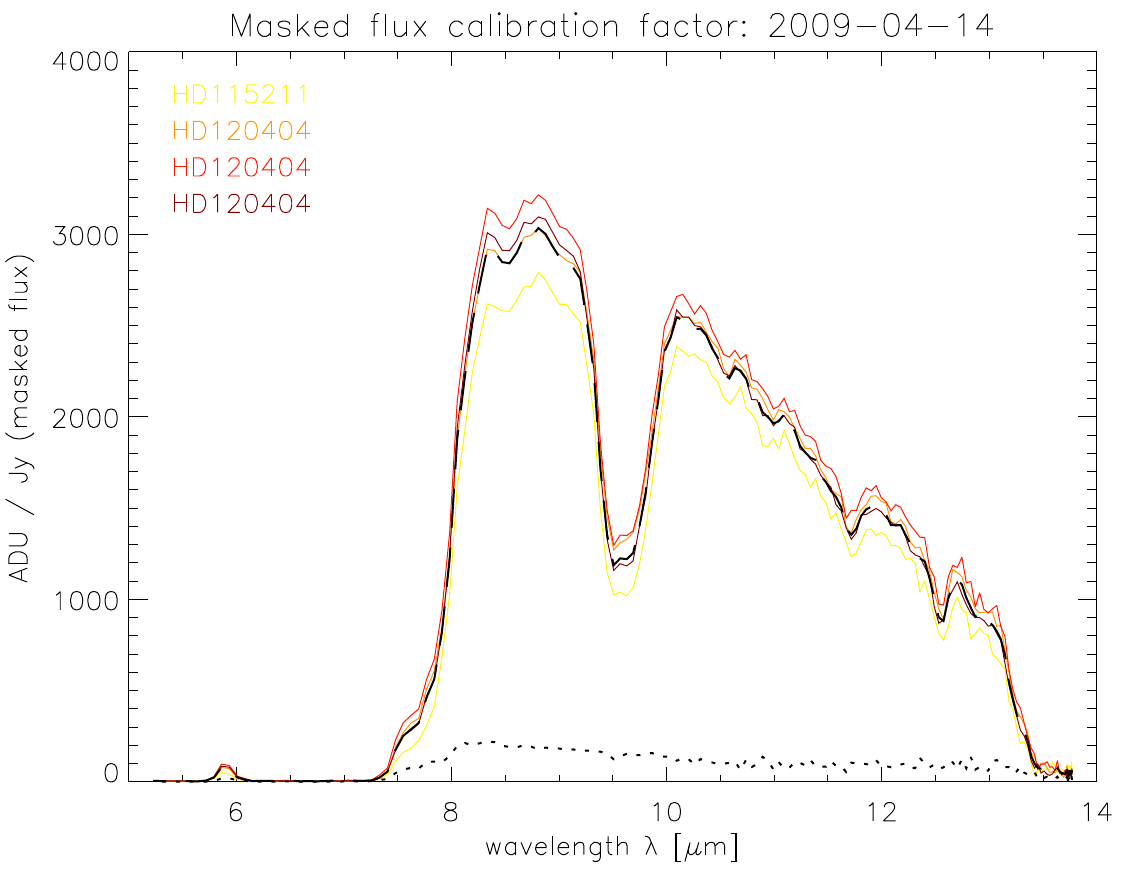}
\par\end{centering}

\caption{Transfer functions for the masked flux on 2008-04-17 and 2009-04-14.\label{fig:transfer}}
\end{figure}

Actually, any atmospheric and instrumental influences on the measured
data should be removed by the calibration process. The data on 2008-04-17
and on 2009-04-14 were calibrated using calibrator stars, which were
observed interleaved with the science data (see Figure~\ref{fig:ambientheader},
calibrators marked in green). With a few exceptions, the calibrator
observed closest in time and airmass was used to calibrate the Circinus
data. In Figure~\ref{fig:transfer}, the transfer functions for the
masked flux on 2008-04-17 and 2009-04-14, calculated from the calibrator
observations, are plotted.
 Both nights seem to have been stable nights and the scatter in the
transfer functions (dotted lines in Figure~\ref{fig:transfer}) is
less than $10\,\%$. This is despite the individual calibrators having
been observed with airmasses between 1.4 and 2.2. It is also worth
mentioning that the largest deviations from the mean transfer function
(black dashed lines in Figure~\ref{fig:transfer}) are observations
of stars other than HD~120404, which is the standard calibrator used
for the Circinus galaxy. This probably means that some of the variations
in the transfer function are due to the uncertainty of the stellar
spectra (taken from the data base by Roy van Boekel) and not so much
due to variations of the atmospheric transparency. On 2009-04-14 only
the three observations of HD~120404 were used for the calibration
because HD~115211 was observed at a very high airmass (c.f. Figure~\ref{fig:ambientheader})
and a few hours before the first observations of the Circinus nucleus
succeeded.

\textbf{The stability of the calibrator transfer functions argue against
the drift at short wavelengths being simply explained by calibration
errors}: the drift is not seen in the raw calibrator data, it comes
from the raw data of the Circinus galaxy alone.

The general offset in flux values in 2009 could, however, be caused
by a variation of the flux of the calibrator, which would have had
to decrease from 2008-04-17 to 2009-04-14. However, HD~120404 is
not listed as a variable star.

In 2008, HD~115211 had a higher transfer function than HD~120404,
while in 2009 it had a lower transfer function (see Figure~\ref{fig:transfer}).
This also argues against a decrease of the flux of HD~120404 from
2008 to 2009, unless HD~115211 underwent an even larger decrease.
The strong relative change in the transfer function of HD~115211
with respect to that of HD~120404 is best explained by the high airmass
($\sim2.05$, see Figure~\ref{fig:ambientheader}) in conjunction
with a lower atmospheric transparency (see Section~\ref{sec:transparency})
in 2009.

\begin{figure}
\centering{}%
\begin{minipage}[t]{0.6\columnwidth}%
\begin{center}
\includegraphics[viewport=140bp 297bp 467bp 552bp,clip,scale=0.75]{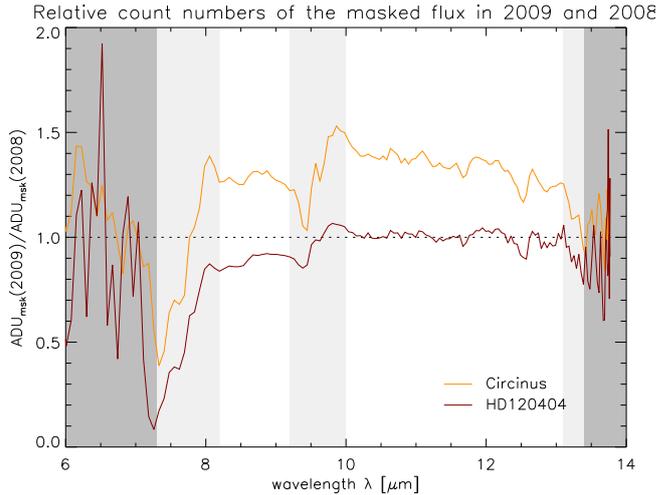}
\par\end{center}%
\end{minipage}%
\begin{minipage}[b][1\totalheight][t]{0.4\columnwidth}%
\caption{Raw counts of the masked flux on 2009-04-14 divided by the raw counts
of the masked flux on 2008-04-17 for both the science target (the
Circinus nucleus) and the calibrator (HD~120404). For the calculation,
the average of all measurements of the masked flux for each epoch
were taken. Areas of lower atmospheric transmission are shaded in
light grey, areas with definitively no transmission in dark grey.\label{fig:relcounts}}
\end{minipage}
\end{figure}

From Figure~\ref{fig:transfer} it appears that the absolute count
rates for HD~120404 on 2009-04-14 are lower than those on 2008-04-17
especially at short wavelengths. A more accurate measure of the difference
in raw count rates is obtained when dividing the (averaged) raw counts
on 2009-04-14 by those on 2008-04-17. This is shown for the masked
flux in Figure~\ref{fig:relcounts} for both HD~120404 and the Circinus
nucleus.
 It becomes clear that the raw count rates of the calibrator only
decreased shortward of $9.5\,\mathrm{\mu m}$; between $9.5$ and
$13.0\,\mathrm{\mu m}$ the ratio of the count rates is consistent
with 1, i.e.\ no change in the absolute flux of the calibrator. Figure~\ref{fig:relcounts}
also shows that it is the raw counts of the Circinus nucleus that
actually increased by roughly $30\,\%$. I therefore conclude that
the count rates of the calibrators are consistent with each other
in the two epochs, taking into account variations the atmospheric
transparency (see Section~\ref{sec:transparency}). \textbf{The offset
in the masked and total flux in Circinus in 2009 is thus not caused
by variability of the calibrator.}

\section{Change in the atmospheric transparency (drift)\label{sec:transparency}}

While the count rates of HD~120404 have remained constant between
2008 and 2009 for $9.5\,\mathrm{\mu m}<\lambda<13.0\,\mathrm{\mu m}$,
there is a $10$ to $15\,\%$ decrease of the count rates of the calibrator
between $8.0$ and $9.5\,\mathrm{\mu m}$ and possibly for $\lambda>13.0\,\mathrm{\mu m}$.
Between $7.0$ and $8.0\,\mathrm{\mu m}$ the count rates in 2009
are only a small fraction of those in 2008. For example the ``spike''
at $7.5\,\mathrm{\mu m}$, visible in the transfer functions in 2008,
is not present in the transfer functions in 2009 (c.f. Figure~\ref{fig:transfer}).
This effect is most likely caused by a different atmospheric transmission
during the two nights. The short wavelength cut-off of the N-band
is caused mainly by absorption by water with some additional absorption
by methane and $\mathrm{N}_{2}\mathrm{O}$ between $7.7$ and $8.0\,\mathrm{\mu m}$.
By comparing the optical thickness of the different species%
\footnote{I could only do a visible comparison as the only site where I could
find the individual species involved in the atmospheric absorption
broken down is at \url{http://www-atm.physics.ox.ac.uk/group/mipas/atlas/}.%
} it is not entirely clear which one of them causes the ``spike''
at $7.5\,\mathrm{\mu m}$, it could be either methane or water.

According to the ESO Ambient Conditions Database%
\footnote{\url{http://archive.eso.org/asm/ambient-server}%
}, the humidity on 2008-04-17 was higher ($20-35\,\%$) than on 2009-04-14
($10-15\,\%$), exactly the opposite than to be expected from the
count rates, which suggest a higher water column in 2009. According
to satellite data%
\footnote{\url{http://www.eso.org/gen-fac/pubs/astclim/forecast/meteo/ERASMUS/par_fapp.txt}%
}, however, the precipitable water vapour in the atmospheric column
above the observatory was $(2.7\pm0.7)\,\mathrm{mm}$ in 2008 and
$(3.3\pm0.7)\,\mathrm{mm}$ in 2009. Although not significant, this
could indicate that indeed a higher water column could be the reason
for the lower count rates at the short wavelength end of the N band
in 2009. Note however that the transfer functions of the calibrator
remained very stable during each night, indicating that the atmospheric
transparency did not change a lot in a single night. \textbf{The atmospheric
transmission is hence not the cause of the nightly drifts in the fluxes.}

The wavelength dependent shape of the ratio of the count rates of
the science source (ignoring the general offset) in Figure~\ref{fig:relcounts}
are similar to that of the calibrator. This means that the atmospheric
conditions for the science source and the calibrators were similar
and that the wavelength dependent changes should calibrate away. What
remains is the directly visible increase in count rates for the Circinus
nucleus in 2009. This is further discussed in Sections~\ref{sec:acquisition}
and \ref{sec:variability-of-Circinus}.

\section{Rotation of the slit on an extended source (drift)\label{sec:rotation}}

\begin{figure}
\begin{centering}
\includegraphics[viewport=140bp 296bp 467bp 552bp,clip,width=0.49\textwidth]{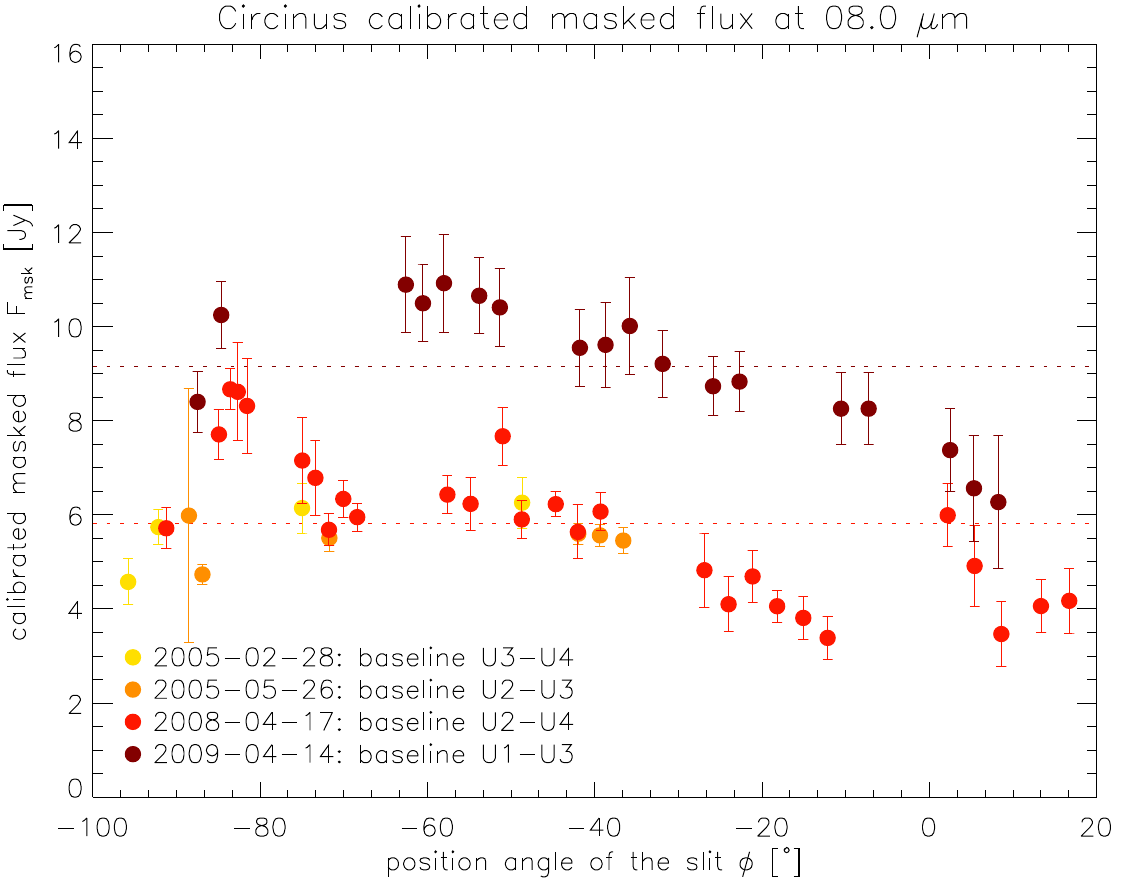}\hfill{}\includegraphics[viewport=135bp 296bp 467bp 552bp,clip,width=0.49\textwidth]{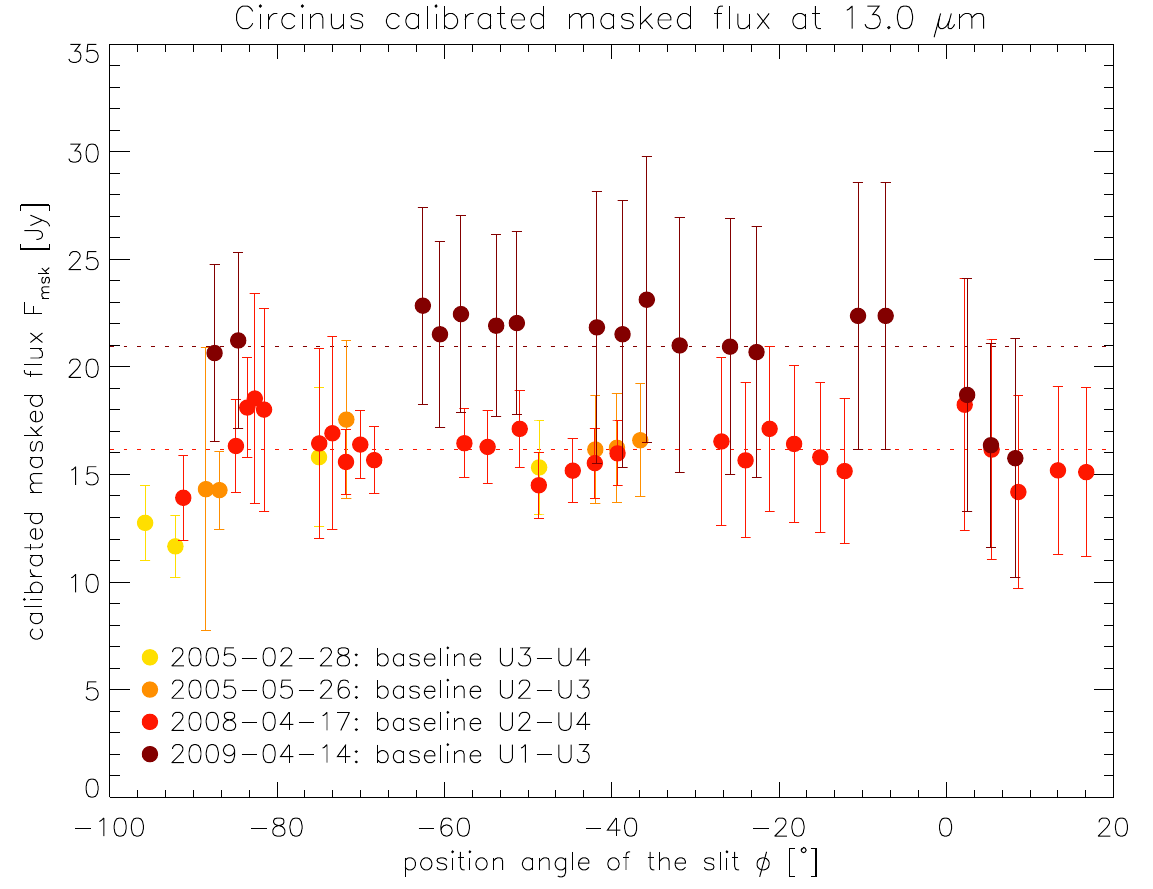}
\par\end{centering}

\caption{Masked flux at $8$ and $10\,\mathrm{\mu m}$ as a function of the
orientation of the slit on the sky for four different baselines and
dates. The average flux values in 2008-04-17 (U2-U4) and 2009-04-14
(U1-U3) are indicated by dashed lines.\label{fig:fluxes_slitangle}}
\end{figure}
\begin{figure}[!b]
\begin{centering}
\includegraphics[viewport=151bp 298bp 437bp 548bp,clip]{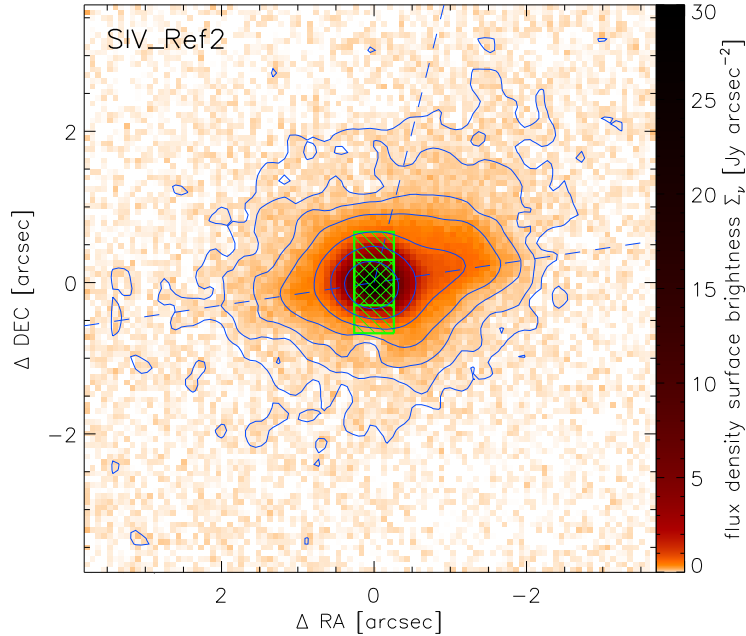}
\par\end{centering}

\caption{VISIR image of the nucleus of the Circinus galaxy at $10.8\,\mathrm{\mu m}$
(logarithmic colour scaling). Overplotted in green are the apertures
for the total (hatched) and masked (cross-hatched) flux for a slit
orientation of $\phi=0{^\circ}$ (North-South).\label{fig:visirimg}}
\end{figure}
\begin{figure}
\centering{}%
\begin{minipage}[t]{0.6\columnwidth}%
\begin{center}
\includegraphics[viewport=140bp 297bp 467bp 552bp,clip,scale=0.75]{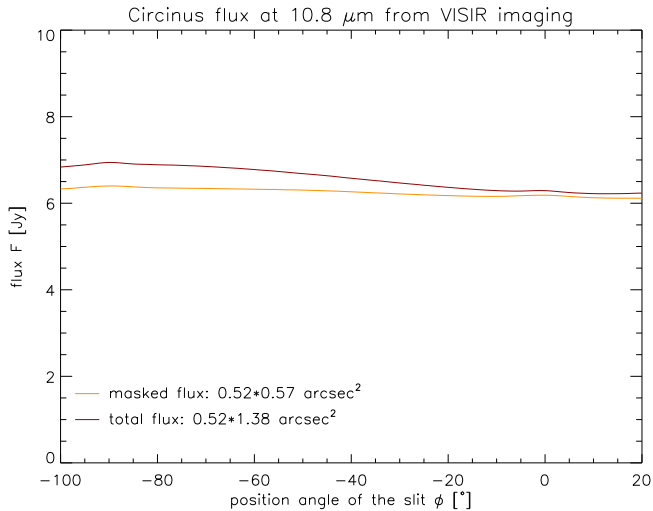}
\par\end{center}%
\end{minipage}%
\begin{minipage}[b][1\totalheight][t]{0.4\columnwidth}%
\caption{Fluxes extracted from the VISIR image at $10.8\,\mathrm{\mu m}$ using
the apertures from 2005-05-26. The fluxes are plotted as a function
of the position angle of the slit like in Figure~\ref{fig:fluxes_slitangle}.
The dependence of the total flux on the slit orientation is larger
than that for the masked flux. \label{fig:visirflx}}
\end{minipage}
\end{figure}
 An explanation for the drift of the fluxes might be that the slit
in MIDI rotates on the extended MIR emission of Circinus. The masked
flux as a function of the orientation of the slit, $\phi$, is shown
in Figure~\ref{fig:fluxes_slitangle}. The figure is very similar
to Figure~\ref{fig:fluxes_hourangle}, because there is almost a
linear relationship between the hour angle and the slit orientation. 

As can be seen from the VISIR image of Circinus at $10.8\,\mathrm{\mu m}$
in Figure~\ref{fig:visirimg}, the nucleus is significantly extended,
when deep images are taken. The VISIR image is seeing limited with
$\mathrm{FWHM}_{\mathrm{nucleus}}(10.8\,\mathrm{\mu m})=400\,\mathrm{mas}$
(i.e. it is larger than the corresponding diffraction limit $\lambda/D=10.8\,\mathrm{\mu m}/8.2\,\mathrm{m}=270\,\mathrm{mas}$
and the calibrator PSF, $\mathrm{FWHM}_{\mathrm{calib}}(10.8\,\mathrm{\mu m})=320\,\mathrm{mas}$).
From the MIDI observations with the ATs it is also clear, that the
source is elongated along $\mathrm{PA}\sim100^{\circ}$ down to spacial
scales of about $100\,\mathrm{mas}$. The average apertures of the
total and masked fluxes as specified in Table~\ref{tab:properties}
are shown in Figure~\ref{fig:visirimg} in green. In contrast to
the rectangular aperture of the total flux, the aperture of the masked
flux is almost quadratic and hence it should be much less affected
by the elongation of the source. In the measured data, this is however
not the case: the dependence of the masked flux on the hour angle
(and similarly on the position angle of the slit) is as large as that
of the total flux (c.f. Figures~\ref{fig:fluxes_hourangle} and \ref{fig:fluxes_slitangle}).
Also, the source is similarly extended at $\lambda>9\,\mathrm{\mu m}$
and the lack of any angle dependence of the flux at long wavelengths
cannot be simply explained by the larger size of the diffraction limit
with respect to $8\,\mathrm{\mu m}$: there should be still be some
slit angle dependence also at the longest wavelengths.

To test this further, I extracted the fluxes at $10.8\,\mathrm{\mu m}$
from the VISIR image using the two apertures from 2005-05-26 (considering
them as some sort of median of the apertures) and for varying position
angles of the slit. Losses due to the relatively small apertures with
respect to the larger PSF size in VISIR (no AO correction) were corrected.
The result is shown in Figure~\ref{fig:visirflx}. As expected, the
flux is much less dependent on the slit position for the masked flux
than for the total flux. In addition, even for the total flux, the
dependency is much weaker than in the measured data. This may be partially
explained by the larger PSF in the VISIR data with respect to the
AO corrected PSF in MIDI ($400\,\mathrm{mas}$ versus $350\,\mathrm{mas}$).
But even when increasing the aperture sizes for the extraction, so
that PSF effects are less relevant, the dependency on the slit angle
remains less than $20\,\%$. On 2008-04-17 the flux at $8.0\,\mathrm{\mu m}$
changed by more than a factor of 2 between $\phi=-85^{\circ}$ and
$\phi=-15^{\circ}$. \textbf{Therefore I conclude that also the slit
orientation cannot be the main reason for the continuous change in
the masked and the total flux.}

In this context it is interesting to note that in the MIDI acquisition
images in 2008 and 2009 $\mathrm{FWHM}\sim(250\pm20)\,\mathrm{mas}$
($\lambda/D=8.8\,\mathrm{\mu m}/8.2\,\mathrm{m}=220\,\mathrm{mas}$)
for the calibrator with the N8.7 filter and $\mathrm{FWHM}\sim(350\pm30)\,\mathrm{mas}$
($\lambda/D=11.8\,\mathrm{\mu m}/8.2\,\mathrm{m}=300\,\mathrm{mas}$)
for the Circinus nucleus and the SiV filter. This is consistent with
the sizes of the PSF derived from the spectra in Section~\ref{sec:PSFsize}.
As far as can be judged from the acquisition images, the PSF for the
Circinus galaxy is not much more elongated than that of the calibrator
stars. Due to the different filters used, it is impossible to estimate
from these numbers to what degree the emission in Circinus is extended
with respect to the PSF of the calibrators like it was done for the
spectra in Section~\ref{sec:PSFsize}.

\section{Polarisation (drift)\label{sec:polarisation}}

If the light of Circinus or the calibrator were polarised to a certain
degree, the rotation of the field of view together with the MIDI and
VLTI optics may lead to a smooth change of the flux with position
angle. However, the degree of polarisation would have to be up to
$50\,\%$ to lead to a change of the flux of up to a factor of 2 (masked
flux on baseline U2-U4 on 2008-04-17 between $\mathrm{HA}=-\textrm{2:30}$
and $\mathrm{HA}=+\textrm{3:00}$). I assume the emission of the calibrator
stars are essentially unpolarised, which is corroborated by the fact
that the transfer functions only show a small scatter and that there
is no continuous change in the transfer function over the night (see
Figure~\ref{fig:transfer}). By consequence, the effect would have
to come from the Circinus nucleus alone. There are no MIR polarisation
measurements for the Circinus galaxy; in the K band the nucleus of
the Circinus galaxy has a polarisation on the order of $3$ to $4\,\%$
\citep{2000Alexander}. The degree of polarisation of NGC 1068 in
the MIR is less than $3\,\%$ \citep{2000Smith,2007Packham}, that
of Mrk~231, a Seyfert 1 galaxy, $8\,\%$ \citep{2001Siebenmorgen}.
It thus seems unlikely that the MIR emission of the Circinus nucleus
is much higher polarised and that the position angle dependent change
of the flux is due to polarisation. Furthermore, the effect should
be equally strong at $8$ and $13\,\mathrm{\mu m}$, unless Circinus
is only polarised at the short wavelength end of the N band. \textbf{I
therefore conclude that it is very unlikely that polarisation is the
reason for the drift}. Nevertheless it remains to be answered, how
the VLTI and MIDI optics react to a polarised source for a rotating
field of view.

\section{Fluxes in the acquisition (offset)\label{sec:acquisition}}

\begin{figure}
\includegraphics[viewport=170bp 344bp 404bp 504bp,clip,scale=0.97]{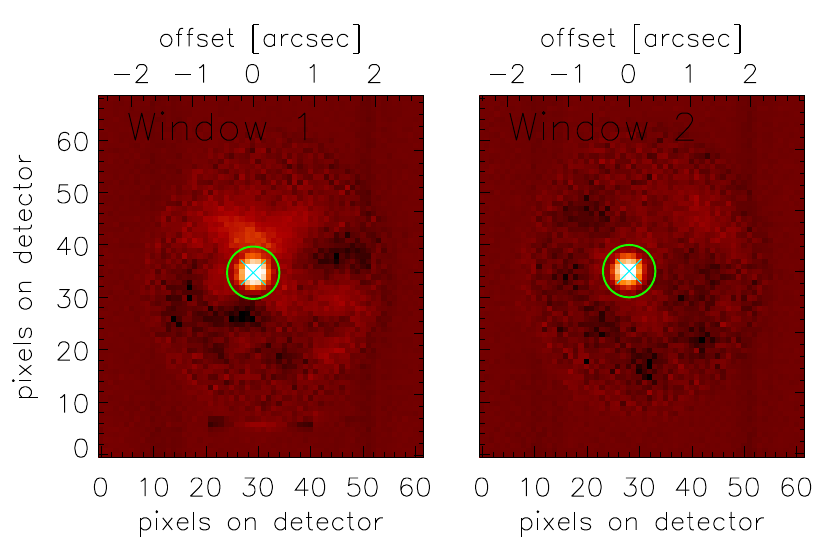}\hfill{}\includegraphics[viewport=197bp 344bp 430bp 504bp,clip,scale=0.97]{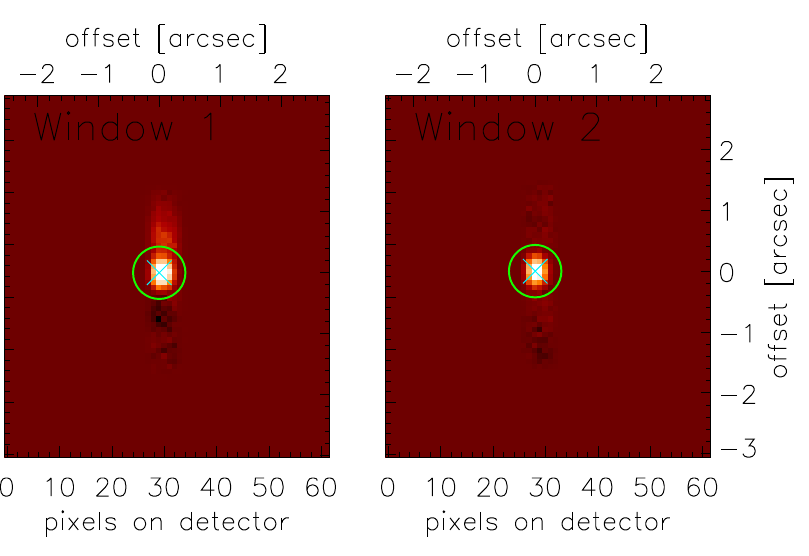}

\caption{Acquisition images (windows 1 and 2, corresponding to beam A and beam
B) of the Circinus galaxy obtained on 2008-04-18 at 03:43. The images
are shown before (left) and after (right) multiplication with the
slit transfer function. The reference pixel for 2008-04-17 is marked
by the blue cross, the aperture used for the photometry indicated
by the green circle.\label{fig:acqimage}}
\end{figure}

The acquisition images can also be used for aperture photometry. An
example of a set of acquisition images is shown in Figure~\ref{fig:acqimage}.
 An aperture of $10\,\mathrm{pixels}$, i.e. $860\,\mathrm{mas}$,
was used for the aperture photometry and is marked by a green circle
in Figure~\ref{fig:acqimage}. The photometry was obtained not only
using the full acquisition images but also a second time, after multiplying
the images with the slit transfer function. The slit transfer function
was directly determined from acquisition images taken with the slit
inserted (c.f. my memo from 2008/12/22).
\begin{figure}
\begin{centering}
\includegraphics[viewport=140bp 296bp 467bp 552bp,clip,width=0.49\textwidth]{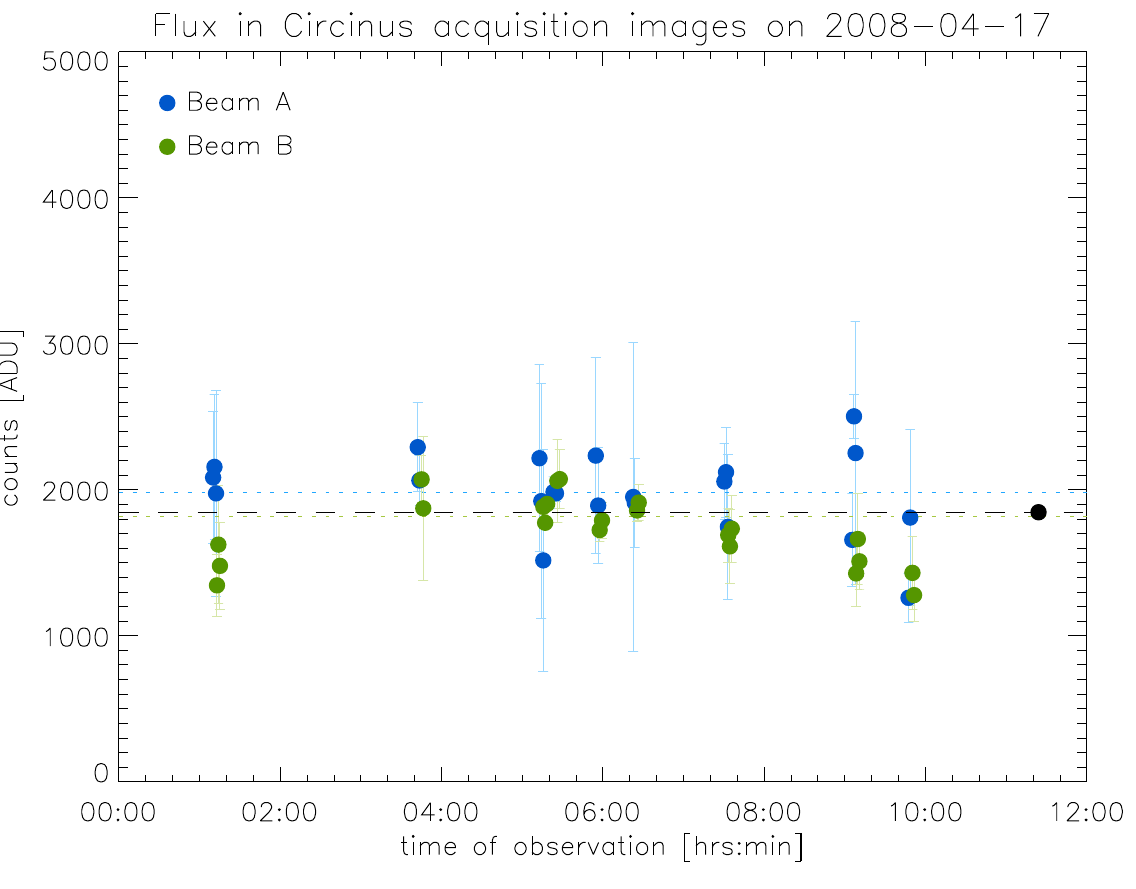}\hfill{}\includegraphics[viewport=135bp 296bp 467bp 552bp,clip,width=0.49\textwidth]{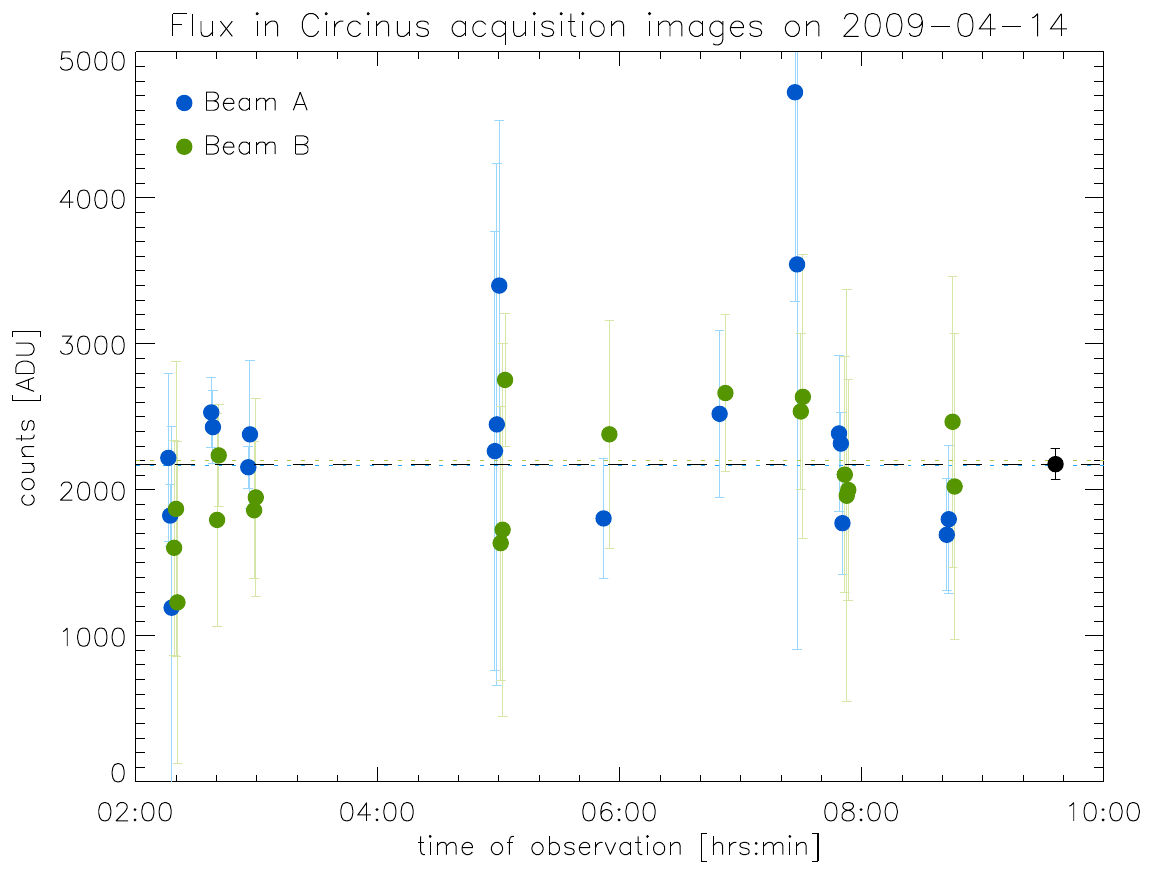}
\par\end{centering}

\begin{centering}
\includegraphics[viewport=140bp 296bp 467bp 552bp,clip,width=0.49\textwidth]{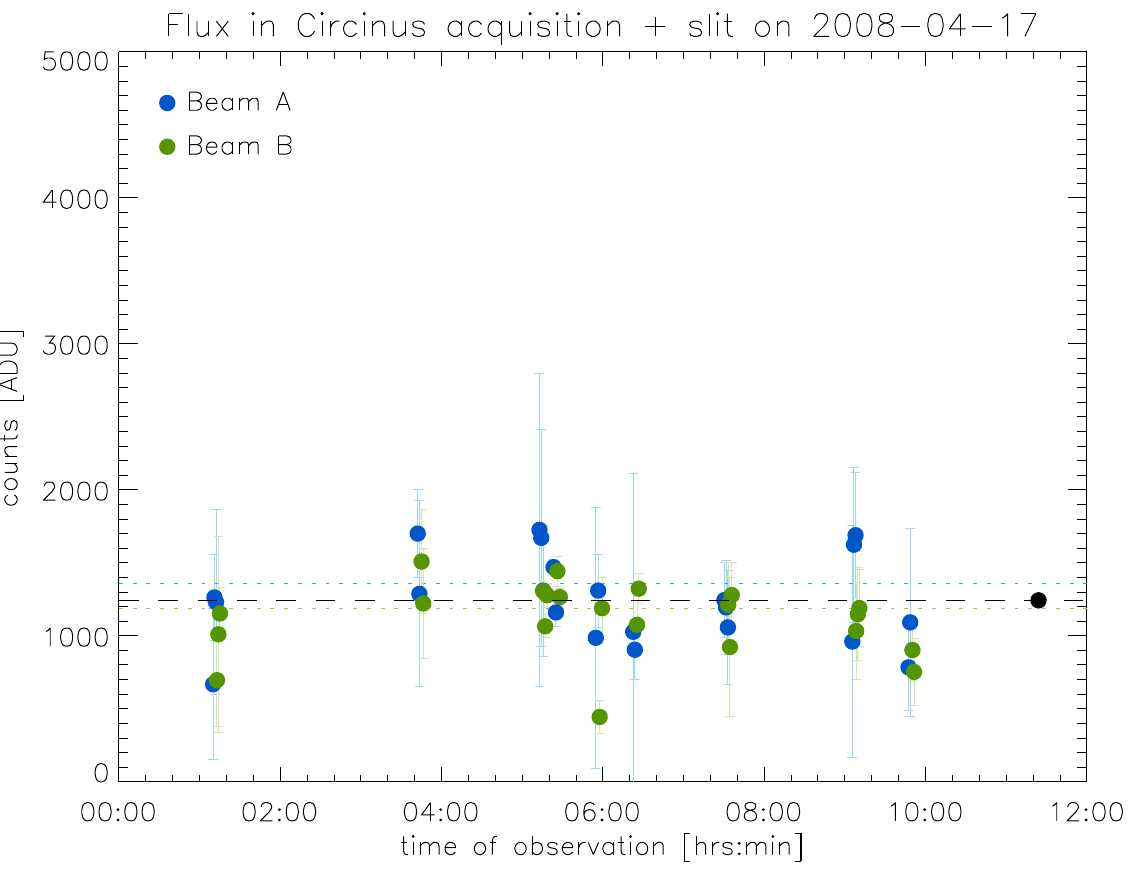}\hfill{}\includegraphics[viewport=135bp 296bp 467bp 552bp,clip,width=0.49\textwidth]{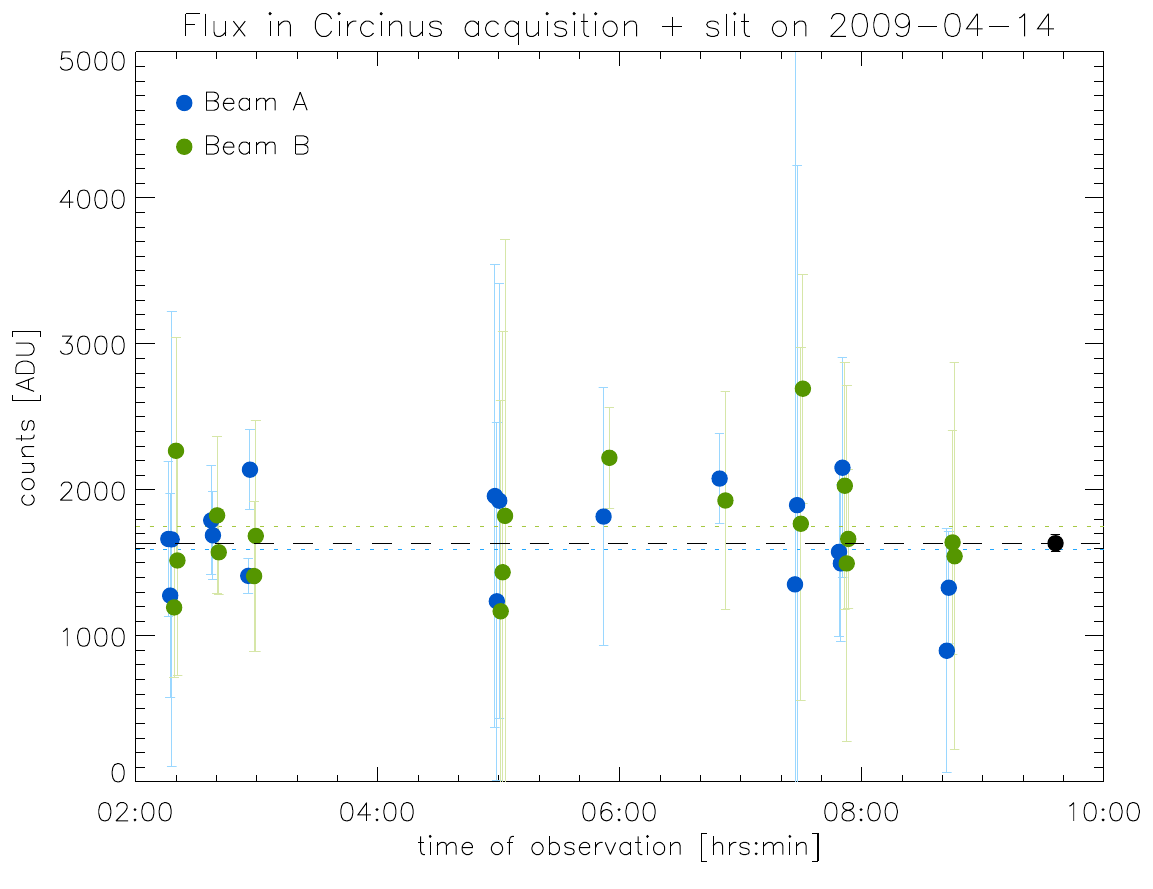}
\par\end{centering}

\caption{Raw count rates of Circinus in the acquisition images on 2008-04-17
(U2-U4, left column) and 2009-04-14 (U1-U3, right column). The fluxes
from beam A are plotted in blue, those from beam B in green. In the
top row the fluxes were directly extracted from the acquisition images.
In the bottom row the acquisition images were multiplied by the slit
transfer function and then the aperture photometry was carried out.
The average count rates are indicated by the dashed lines and the
black points and they are given in Table~\ref{tab:acqfluxes}.\label{fig:acqflux-cir}}
\end{figure}
\begin{figure}
\begin{centering}
\includegraphics[viewport=140bp 296bp 467bp 552bp,clip,width=0.49\textwidth]{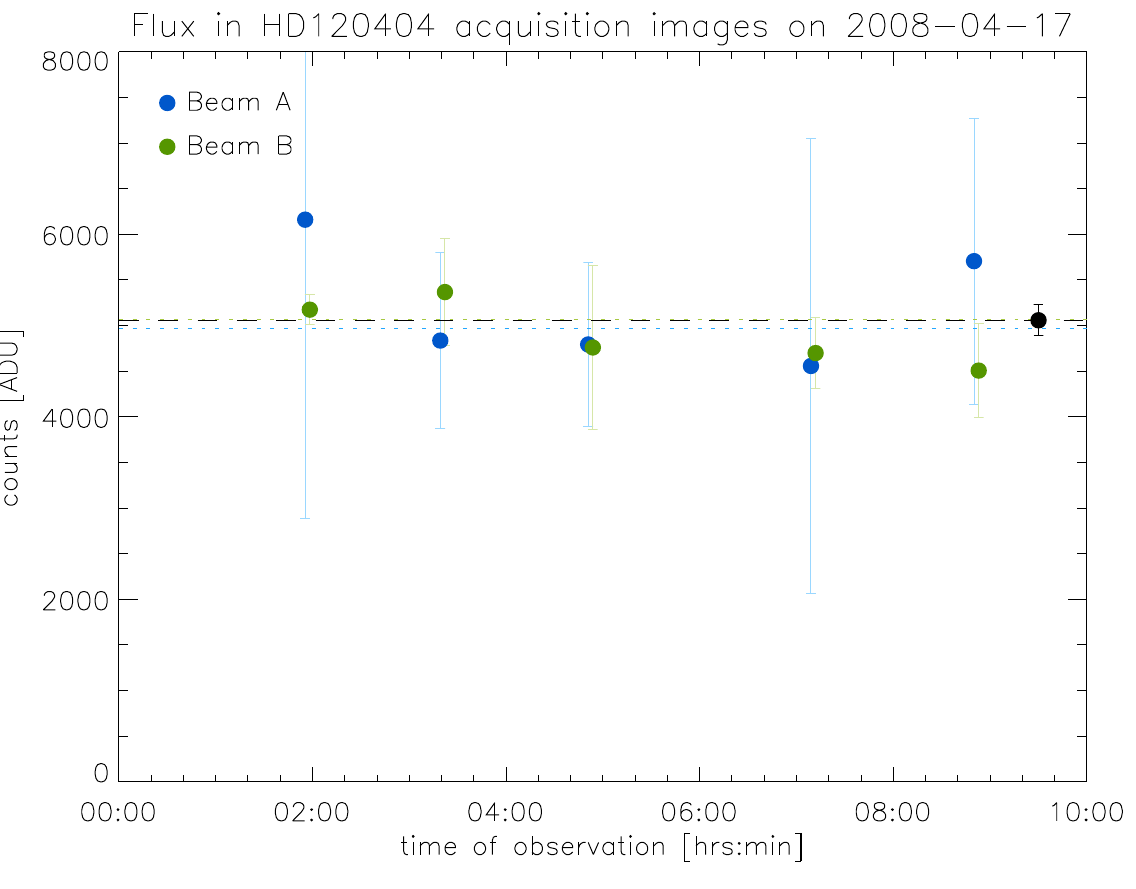}\hfill{}\includegraphics[viewport=135bp 296bp 467bp 552bp,clip,width=0.49\textwidth]{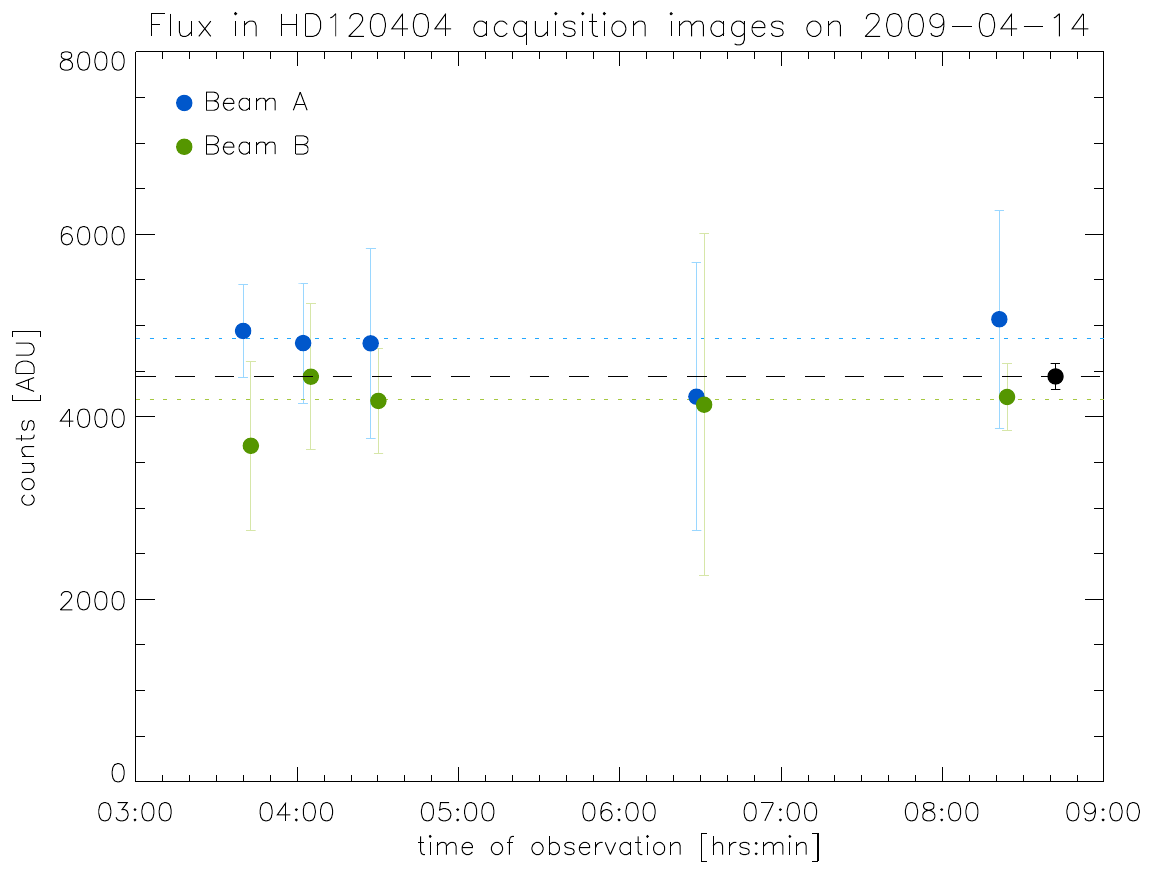}
\par\end{centering}

\begin{centering}
\includegraphics[viewport=140bp 296bp 467bp 552bp,clip,width=0.49\textwidth]{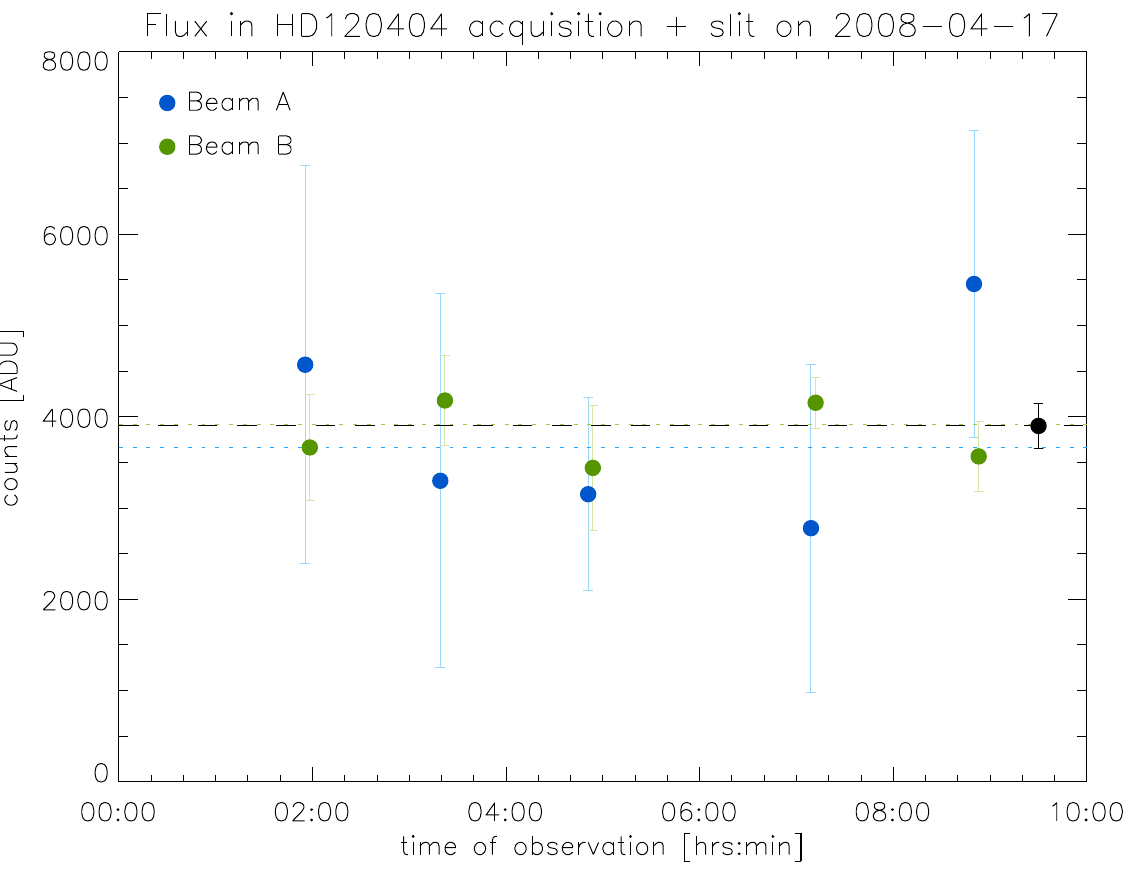}\hfill{}\includegraphics[viewport=135bp 296bp 467bp 552bp,clip,width=0.49\textwidth]{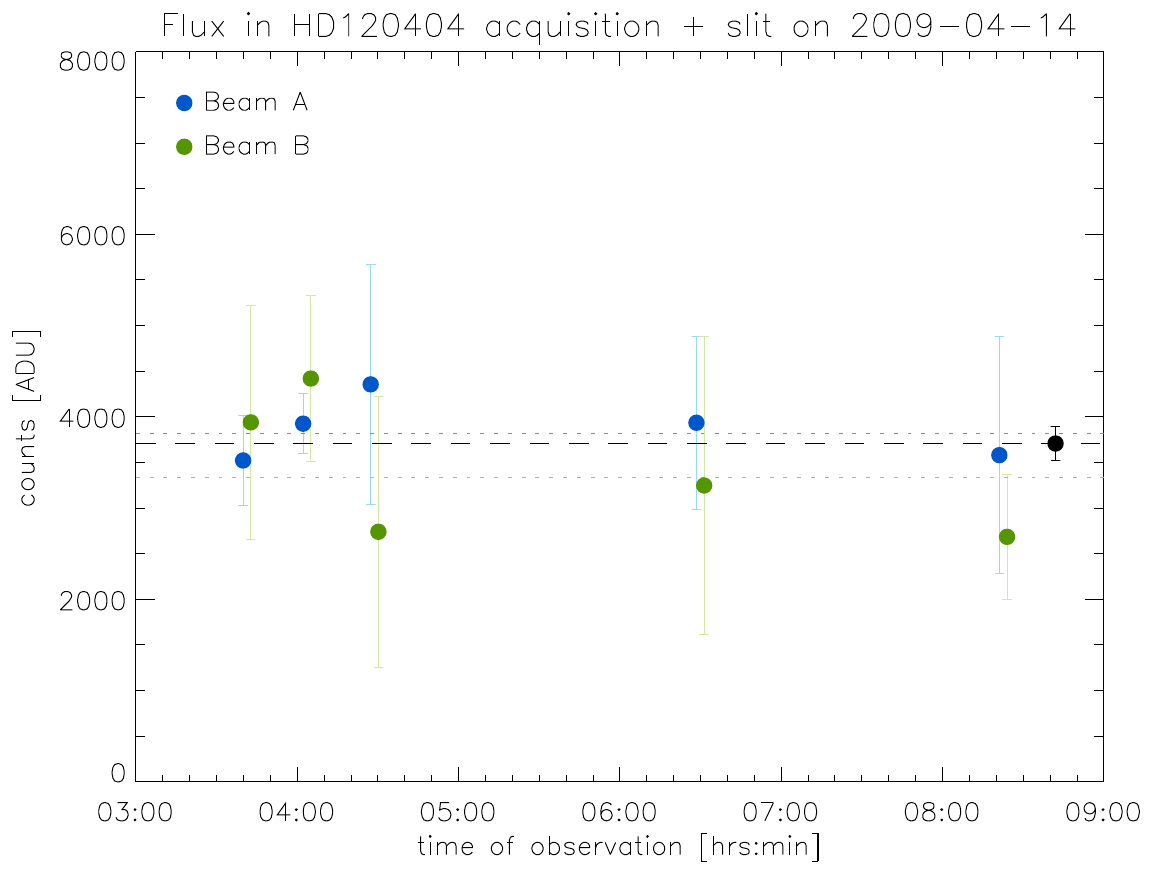}
\par\end{centering}

\caption{Same as Figure~\ref{fig:acqflux-cir} but for the calibrator HD~120404.
\label{fig:acqflux-cal}}
\end{figure}

The raw count rates of the acquisition images obtained with MIDI on
2008-04-17 and 2009-04-14 are plotted in Figures~\ref{fig:acqflux-cir}
and \ref{fig:acqflux-cal} for the Circinus nucleus and HD~120404
respectively. For Circinus mainly the SiC filter was used, while for
HD~120404 the N8.7 filter was used. Due to the different filters
it is not possible to calibrate the photometry for Circinus, only
the raw count rates can be compared. The average values for each night
are indicated by the black dashed line and the black data point at
the end of the observing period. They are summarised in Table~\ref{tab:acqfluxes}.
\begin{table}[b]
\begin{tabular}{cc@{\extracolsep{1pt}}rr@{ }r@{ }r@{\extracolsep{1pt}}rr@{ }r@{ }r}
\hline 
object & slit &  & \multicolumn{3}{c}{2008-04-17} &  & \multicolumn{3}{c}{2009-04-14}\tabularnewline
\hline 
Circinus & no &  & $1849$ & $\pm$ & $47$ &  & $2180$ & $\pm$ & $121$\tabularnewline
(SiC) & yes &  & $1240$ & $\pm$ & $46$ &  & $1643$ & $\pm$ & $64$\tabularnewline
\multicolumn{2}{r}{slit losses} &  & \multicolumn{3}{r}{$(33\pm4)\,\%$} &  & \multicolumn{3}{r}{$(25\pm6)\,\%$}\tabularnewline
 &  &  &  &  &  &  &  &  & \tabularnewline
HD~120404 & no &  & $5058$ & $\pm$ & $172$ &  & $4443$ & $\pm$ & $140$\tabularnewline
(N8.7) & yes &  & $3900$ & $\pm$ & $248$ &  & $3708$ & $\pm$ & $191$\tabularnewline
\multicolumn{2}{r}{slit losses} &  & \multicolumn{3}{r}{$(23\pm6)\,\%$} &  & \multicolumn{3}{r}{$(17\pm6)\,\%$}\tabularnewline
\hline 
\end{tabular}\hfill{}%
\begin{tabular}{ccc@{ }r@{ }r@{ }r@{\extracolsep{1pt}}rc@{ }r@{ }r@{ }r}
\hline 
object & filter & \multicolumn{4}{c}{2008-04-17} &  & \multicolumn{4}{c}{2009-04-14}\tabularnewline
\hline 
Circinus & SiC & $365$ & $\pm$ & $3$ & mas &  & $340$ & $\pm$ & $4$ & mas\tabularnewline
 &  &  &  &  &  &  &  &  &  & \tabularnewline
 &  &  &  &  &  &  &  &  &  & \tabularnewline
HD~120404 & N8.7 & $247$ & $\pm$ & $4$ & mas &  & $237$ & $\pm$ & $4$ & mas\tabularnewline
 &  &  &  &  &  &  &  &  &  & \tabularnewline
\hline 
\end{tabular}

\caption{Average count rates (left table, in ADU) and PSF sizes (right table)
determined from the acquisition images. \label{tab:acqfluxes}}
\end{table}

In all these plots, no trend of the flux with time of observation
within one epoch is discernible. This is consistent with the flux
values of Circinus to only change at short wavelengths, because the
SiC filter with $\lambda_{\mathrm{c}}=11.79$ was used for the Circinus
acquisition. By consequence the acquisition images give no clue, what
may cause the drift of the fluxes at short wavelengths during one
night. 

On the other hand, higher flux rates for Circinus in 2009 than in
2008 were measured in the acquisition images. The difference is not
significant for the individual measurements but it is for the average
with an increase of the average count rate by $(18\pm8)\,\%$. For
HD~120404, the opposite is the case, the average of the count rates
in 2009 is lower by $(12\pm4)\,\%$ than in 2008. Considering that
the N8.7 filter was used for the calibrator this is consistent with
the change of the atmospheric conditions and the $10$ to $15\,\%$
decrease of the atmospheric transparency at short wavelengths already
discussed in Section~\ref{sec:transparency}. Note that due to the
different filters used for Circinus and the calibrator a direct comparison
is not possible. \textbf{In any case, there seems to have been an
increase in the Circinus flux from 2008 to 2009 already in the acquisition
images by about $20$ to $30\,\%$ with respect of that of the calibrator.}

The increase in Circinus is even larger when the acquisition image
is multiplied by the slit transfer function. Then the increase is
$(32\pm9)\,\%$ for the raw counts alone. This enhancement can be
explained be reduced slit losses in 2009. Before January 2009, the
reference pixel for centring the source was not centred in the MIDI
slit (see my memo from 2008/12/22). This can be clearly seen in Figure~\ref{fig:acqimage}:
The blue cross is not centred in the slit. This causes stronger slit
losses. In principle, the calibration should corrected for the slit
losses, however the slit losses are a bit larger for the Circinus
galaxy than for the calibrator because the PSF of the galaxy is slightly
more extended. This could have lead to a further apparent increase
in the Circinus flux from 2008 to 2009. The values for the slit losses
in Table~\ref{tab:acqfluxes} are not directly comparable due to
the different filters used: the slit losses should be larger at longer
wavelengths than at short wavelengths, due to the wavelength dependent
size of the PSF. The increase in flux of Circinus is however not wavelength
dependent (apart from an atmospheric contribution, see Section~\ref{sec:transparency}
and Figure~\ref{fig:relcounts}). \textbf{Thus the adjustment of
the reference pixel in January 2009 can only have lead to a minor
apparent increase in the Circinus flux.} The majority comes from the
increase of the raw count rates for Circinus.

In 2008, one acquisition of Circinus was obtained using the N8.7 filter.
Calibration gives a flux of $(4.92\pm0.43)\,\mathrm{Jy}$ for Circinus
at $\lambda_{\mathrm{c}}=8.64\,\mathrm{\mu m}$. This value fits perfectly
to the average of all total flux measurements with MIDI (see Figure~\ref{fig:circnband}).

\section{Variability of Circinus\label{sec:variability-of-Circinus}}

\begin{figure}
\centering{}%
\begin{minipage}[t]{0.6\columnwidth}%
\begin{center}
\includegraphics[viewport=140bp 297bp 467bp 552bp,clip,scale=0.75]{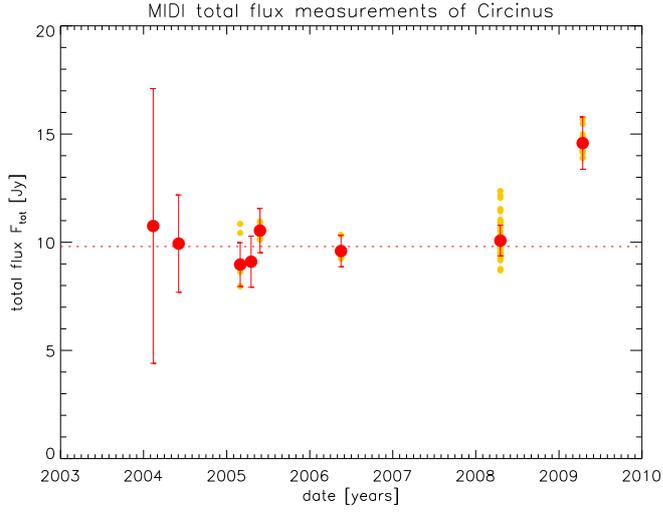}
\par\end{center}%
\end{minipage}%
\begin{minipage}[b][1\totalheight][t]{0.4\columnwidth}%
\caption{Total flux of the Circinus nucleus at $12\,\mathrm{\mu m}$ as a function
of the observing date. The large red points with error bars mark the
weighted average of the individual flux measurements for each of the
epochs. The individual measurements are indicated by the yellow points.
Especially in 2008 and 2009 more than 10 individual spectra were obtained
for each epoch. The average of the measurements from 2004 to 2008
is indicated by the red dotted line. \label{fig:fluxepoch}}
\end{minipage}
\end{figure}

\begin{figure}
\begin{centering}
\includegraphics[viewport=45bp 223bp 552bp 605bp,clip,width=0.9\textwidth]{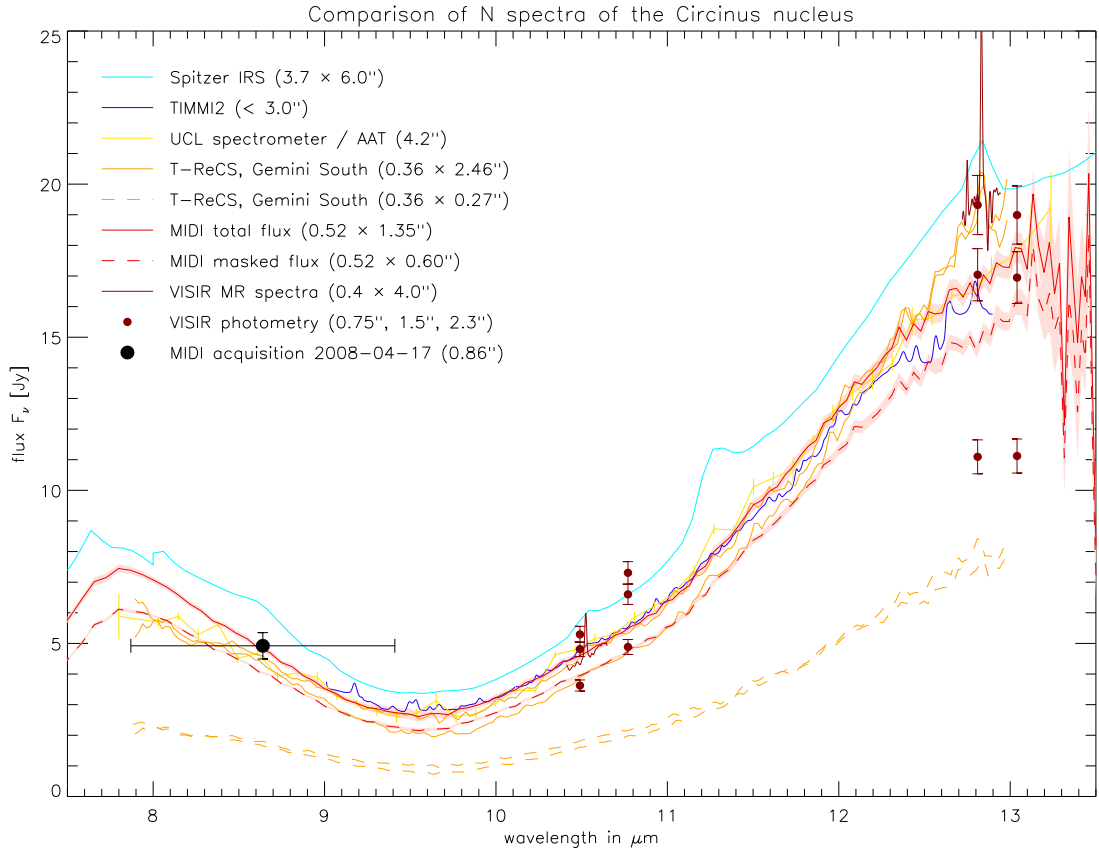}
\par\end{centering}

\caption{Comparison of various spectra and photometric measurements of the
Circinus nucleus in the N band.\label{fig:circnband}}
\end{figure}

The previous section directly leads to the conclusion that the higher
values measured in 2009 are due to a variability of the source itself.
Figure~\ref{fig:fluxepoch} shows the fluxes measured by MIDI at
$12\,\mathrm{\mu m}$ as a function of the observing date. Clearly
the increase in the flux by almost $50\,\%$ is apparent for the measurements
in 2009; all measurements before 2009 seem to be consistent with each
other considering the errors. Also several other MIR measurements
obtained in the last years all agree to the averaged MIDI spectrum
within $30\,\%$ (most of the differences being probably due to differences
in aperture, see Figure~\ref{fig:circnband}).

\begin{figure}
\centering{}%
\begin{minipage}[t]{0.6\columnwidth}%
\begin{center}
\includegraphics[viewport=145bp 297bp 467bp 552bp,clip,scale=0.75]{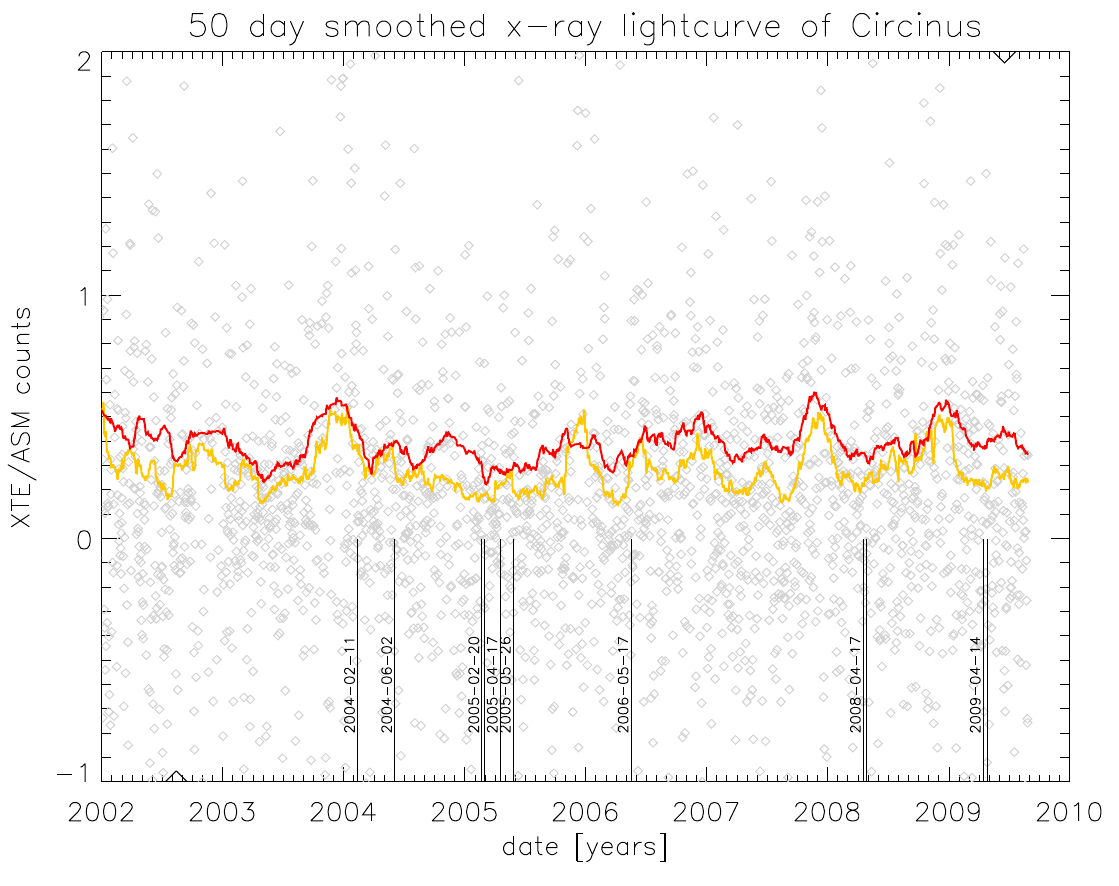}
\par\end{center}%
\end{minipage}%
\begin{minipage}[b][1\totalheight][t]{0.4\columnwidth}%
\caption{X-ray light curve from RXTE ASM data.\label{fig:xraylightcurve}}
\end{minipage}
\end{figure}

There are no studies of the variability of the Circinus nucleus in
the infrared (i.e. for neither the NIR nor the MIR). In the X-ray
regime, the nucleus of Circinus ``is observed as consistently not
variable'' over a time span of about 9 years \citep{2009Winter}.
Hard X-ray observations show no clear indications for variability
of the nucleus either, although the results are inconclusive \citep{2009Yang}.
In the X-ray monitoring with RXTE%
\footnote{\url{http://xte.mit.edu/asmlc/ASM.html}%
}, no strong increase of the X-ray flux in 2009 is discernible either
(see Figure~\ref{fig:xraylightcurve}).
 Because Circinus is Compton thick with $N_{\mathrm{H}}\sim4\cdot10^{24}$,
one has to be careful to use X-rays as a tracer for the luminosity
of the accretion disk, but for energies above 13 keV the directly
transmitted nuclear emission seems to be visible \citep{2009Yang}.

\begin{figure}
\centering{}%
\begin{minipage}[t]{0.6\columnwidth}%
\begin{center}
\includegraphics[viewport=145bp 297bp 467bp 552bp,clip,scale=0.75]{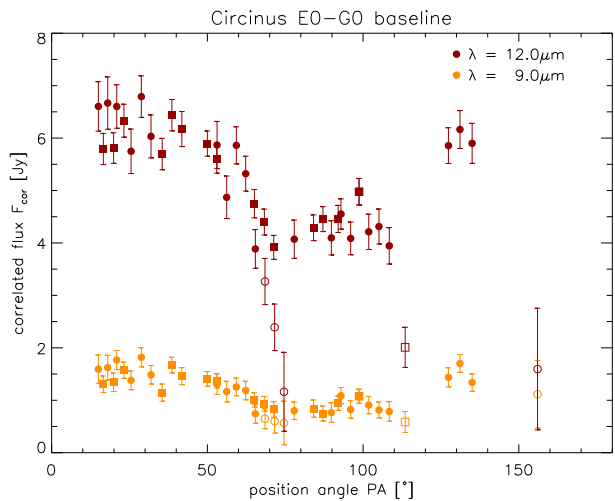}
\par\end{center}%
\end{minipage}%
\begin{minipage}[b][1\totalheight][t]{0.4\columnwidth}%
\caption{Correlated fluxes as a function of the position angle for two wavelengths
on the E0-G0 baseline from two observing dates, 2008-04-25 (circles)
and 2009-04-26 (boxes). Measurements where no fringe track was actually
carried out or which appear otherwise ``dodgy'' are marked by open
symbols.\label{fig:at-fluxes}}
\end{minipage}
\end{figure}

A problem with a possible increase of the intrinsic MIR flux of the
Circinus nucleus is however, that because of the time scale of variation,
this variable emission should come from a region smaller than about
$1\,\mathrm{ly}\approx0.3\,\mathrm{pc}$, which corresponds to $15\,\mathrm{mas}$.
This is just the size of the central disk component probed with MIDI.
Therefore the increase in the flux should, to a large degree, also
be seen in the correlated fluxes. This is, however, not the case:
the absolute increase of the masked flux is on the order of $3$ to
$5\:\mathrm{Jy}$, all the correlated fluxes in 2009 on the U1-U3
baseline have $F_{\mathrm{cor}}<2.0\,\mathrm{Jy}$ with most of them
actually on the order of only $0.5\,\mathrm{Jy}$. They simply cannot
have increased by a few $\mathrm{Jy}$. Furthermore, AT measurements
on the E0-G0 baseline were carried out both on 2008-04-25 and on 2009-04-26,
a few days after the UT observations in these two years. In Figure~\ref{fig:at-fluxes},
the correlated fluxes of these measurements are plotted as a function
of the position angle for two wavelengths, $9$ and $12\,\mathrm{\mu m}$.
 Both observations agree well (almost within $1\sigma$) even though
the night in 2009 (boxes) was less stable than the one in 2008 (circles)
and the transfer function in 2009 varied by up to $50\,\%$ (for a
more detailed discussion see also my memo from 09/06/2009). There
is hence no indication for variability in this AT data, which probes
spacial scales of $\sim80\,\mathrm{mas}$. This leaves three explanations:
\begin{itemize}
\item the variability comes from regions larger than $\sim80\,\mathrm{mas}$
\item there was just an ``outburst'' on 2008-04-14 and the MIR flux reverted
to its normal state by 2009-04-26
\item there was no intrinsic increase in the MIR emission of Circinus. 
\end{itemize}
The first two explanations are very unlikely. \textbf{Thus the most
likely explanation is that there was no increase in the MIR emission
of the Circinus nucleus and that the increased count rates are due
to an instrumental effect.} Note that no useful photometry could be
measured with the ATs.

\section{Conclusion}

The MIDI photometry data of the nucleus of the Circinus galaxy shows
a general \textit{offset} of the flux values for 2009 and a slow \textit{drift}
of the total fluxes at short wavelengths. Several possible reasons
for these variations were investigated, but no conclusive explanation
for either of these variations could be found. The implications of
the ``unexplainable'' variations on the credibility of the data
obtained with MIDI remain to be discussed.

\bibliographystyle{aamod}
\bibliography{circinus_photstab}

\end{document}